\documentclass[onecolumn]{aastex63}
\usepackage{longtable}
\usepackage{booktabs}
\usepackage[section]{placeins}
\usepackage{makecell}
\usepackage{amsmath} 
\usepackage{cases}
\shorttitle{VLA Observations of L1251 VLA 6}
\shortauthors{Nederlander et al.}
\usepackage{graphicx}
\usepackage[flushleft]{threeparttable}
\usepackage{scalerel}
\usepackage{tikz}
\usepackage{comment} 
\usetikzlibrary{svg.path}

\definecolor{orcidlogocol}{HTML}{A6CE39}
\tikzset{
  orcidlogo/.pic={
    \fill[orcidlogocol] svg{M256,128c0,70.7-57.3,128-128,128C57.3,256,0,198.7,0,128C0,57.3,57.3,0,128,0C198.7,0,256,57.3,256,128z};
    \fill[white] svg{M86.3,186.2H70.9V79.1h15.4v48.4V186.2z}
                 svg{M108.9,79.1h41.6c39.6,0,57,28.3,57,53.6c0,27.5-21.5,53.6-56.8,53.6h-41.8V79.1z M124.3,172.4h24.5c34.9,0,42.9-26.5,42.9-39.7c0-21.5-13.7-39.7-43.7-39.7h-23.7V172.4z}
                 svg{M88.7,56.8c0,5.5-4.5,10.1-10.1,10.1c-5.6,0-10.1-4.6-10.1-10.1c0-5.6,4.5-10.1,10.1-10.1C84.2,46.7,88.7,51.3,88.7,56.8z};
  }
}

\newcommand\orcidicon[1]{\href{https://orcid.org/#1}{\mbox{\scalerel*{
\begin{tikzpicture}[yscale=-1,transform shape]
\pic{orcidlogo};
\end{tikzpicture}
}{|}}}}

\usepackage{hyperref} 

\begin{document}
\title{An Outbursting Protostar: The environment of L1251 VLA 6}

\author{Ava Nederlander \orcidicon{0000-0002-4884-665X}}
\affiliation{Astronomy Department and Van Vleck Observatory, Wesleyan University, Middletown, CT 06459, USA}
\affiliation{Department of Electrical and Computer Engineering, Stony Brook University, Stony Brook, NY 11794-3800, USA}

\author{Adele Plunkett \orcidicon{0000-0002-9912-5705}}
\affiliation{National Radio Astronomy Observatory, 520 Edgemont Road Charlottesville, VA 22903, USA}

\author{Antonio Hales \orcidicon{0000-0001-5073-2849}}
\affiliation{National Radio Astronomy Observatory, 520 Edgemont Road Charlottesville, VA 22903, USA}
\affiliation{Joint ALMA Observatory, Avenida Alonso de Córdova 3107, Vitacura 7630355, Santiago, Chile}

\author{\'{A}gnes K\'{o}sp\'{a}l \orcidicon{0000-0001-7157-6275}}
\affiliation{Konkoly Observatory, HUN-REN Research Centre for Astronomy and Earth Sciences, Konkoly Thege Mikl\'{o}s \'{u}t 15-17, H-1121 Budapest, Hungary}
\affiliation{Max Planck Institute for Astronomy, K\"onigstuhl 17, D-69117 Heidelberg, Germany}
\affiliation{ELTE E\"otv\"os Lor\'and University, Institute of Physics, P\'azm\'any P\'eter s\'et\'any 1/A, 1117 Budapest, Hungary}
\affiliation{CSFK, MTA Centre of Excellence, Konkoly-Thege Mikl\'os \'ut 15-17, 1121 Budapest, Hungary}

\author{Jacob A. White \orcidicon{0000-0001-8445-0444}}
\affiliation{National Radio Astronomy Observatory, 520 Edgemont Road Charlottesville, VA 22903, USA}
\affiliation{EarthDaily Analytics, 1055 Canada Pl  \#33, Vancouver, BC V6C 3L5, Canada}

\author{Makoto A. Johnstone \orcidicon{0000-0001-7690-3976}}
\affiliation{Department of Physics, Middlebury College, Middlebury, VT 05753, USA}

\author{M\'{a}ria Kun \orcidicon{0000-0002-7538-5166}}
\affiliation{Konkoly Observatory, HUN-REN Research Centre for Astronomy and Earth Sciences, Konkoly Thege Mikl\'{o}s \'{u}t 15-17, H-1121 Budapest, Hungary}
\affiliation{CSFK, MTA Centre of Excellence, Konkoly-Thege Mikl\'os \'ut 15-17, 1121 Budapest, Hungary}

\author{P\'{e}ter \'{A}brah\'{a}m  \orcidicon{0000-0001-6015-646X}}
\affiliation{Konkoly Observatory, HUN-REN Research Centre for Astronomy and Earth Sciences, Konkoly Thege Mikl\'{o}s \'{u}t 15-17, H-1121 Budapest, Hungary}
\affiliation{ELTE E\"otv\"os Lor\'and University, Institute of Physics, P\'azm\'any P\'eter s\'et\'any 1/A, 1117 Budapest, Hungary}
\affiliation{CSFK, MTA Centre of Excellence, Konkoly-Thege Mikl\'os \'ut 15-17, 1121 Budapest, Hungary}
\affiliation{Department of Astrophysics, University of Vienna, T\'urkenschanzstr 17, A-1180 Vienna, Austria}

\author{Anna G. Hughes \orcidicon{0000-0002-3446-0289}}
\affiliation{Entanglement, Inc., 109 Greene St, New York, NY 10012, USA}

\begin{abstract}
Young protostars that undergo episodic accretion can provide insight into the impact on their circumstellar environments while matter is accreted from the disk onto the protostar.  
IRAS 22343+7501 is a four component protostar system with one of those being a fading outbursting protostar referred to as L1251 VLA 6.  Given the rarity of YSOs undergoing this type of accretion, L1251 VLA 6 can elucidate the fading phase of the post-outburst process.  Here we examine structure in the disk around L1251 VLA 6 at frequencies of 10\,GHz and 33\,GHz with the Karl G. Jansky Very Large Array (VLA). We model the disk structure using Markov chain Monte Carlo (MCMC).  This method is then combined with a parametric ray-tracing code to generate synthetic model images of an axisymmetric disk, allowing us to characterize the radial distribution of dust in the system.  The results of our MCMC fit show that the most probable values for the mass and radius are consistent with values typical of Class I objects.  We find that the total mass of the disk is $0.070^{+0.031}_{-0.2} \rm ~ M_{\sun}$ and investigate the conditions that could cause the accretion outburst.  We conclude that the eruption is not caused by gravitational instability and consider alternative explanations and trigger mechanisms.

\end{abstract}
\section{Introduction}

The star formation process begins in the dense regions of interstellar molecular clouds in which a gravitational collapse causes gas and dust to heat up and begin forming young stellar objects (YSOs).  The rotation of the collapsing cloud core produces an accretion disk, which can add mass to the protostar.  Mass buildup can be accelerated by episodic accretion in YSOs.  Theoretical considerations \citep[e.g.,][]{Vorobyov2015} suggest that Sun-like stars build up their mass during periods of significantly enhanced accretion (i.e., episodic accretion). The most substantial outbursts manifest during the embedded phase, wherein the accretion disk is supplied with gas originating from the collapsing envelope.  Episodic accretion can play a crucial role in the evolution of circumstellar disks, and it is considered to have an impact on planet formation.  Outbursts have been shown to potentially change the chemistry and mineralogy of the surrounding circumstellar disk \citep[][]{abraham2009, rab2017}; could spur the growth of small solids through, e.g., evaporation and recondensation from a rapid evolution of the ice line \citep[][]{cieza2018}; and may lead to the large spread of observed protostellar luminosities \citep[][]{Fischer2023}.  Furthermore, in the common model of star formation, YSOs surrounded by circumstellar and/or circumbinary disks regularly generate strong, episodic jets that enable interpretation of the evolutionary state and activity of those YSOs \citep[][]{Arce2007,Pech2010,Vorobyov2018}.

While it is predicted that all Sun-like stars undergo several outbursts during their pre-main sequence evolution \citep[][]{audard2014}, there are very few detected young stars that currently experience accretion outbursts \citep[][]{Fischer2023}.  This is in large part due to the phenomenon becoming observable only as the circumstellar envelope thins \citep[][]{audard2014}.  Given that there are only a few dozen stars of this sort that are known, it is crucial for our understanding to examine each new item \citep[][]{Semkov2021}. Among those first discovered were the original prototypes FU Orionis and EX Lupi \citep[][]{Herbig1989}.  Subsequent to \citet{audard2014}, even more outbursting sources have been found, as the Gaia satellite has been remarkably successful in discovering new, outbursting sources \citep[e.g.,][]{Hillenbrand2018, Szegedi-Elek2020, Hodapp2019, Hillenbrand2019, Hodapp2020, Miera2022, Nagy2023, Siwak2023}, making this an important phenomenon to study more broadly.  These discoveries have expanded our understanding of this phenomenon, and it is now known that outbursts occur in various YSOs, encompassing different classes and intermediate/ambiguous properties.  Mechanisms of mass accumulation and disk heating, leading to accretion bursts, are discussed in several theoretical papers \citep[e.g.,][]{Bell1994, Bell1995, Vorobyov2006, Zhu2009, Bae2014}. The existence of an envelope is presupposed in most scenarios, therefore FUors may signify the change from embedded to optically visible young stars. In one of the proposed mechanisms, episodic outbursts are caused by matter accumulation in the inner disk region as a result of angular momentum rearrangements in the outer disk caused by gravitational instabilities (GI), as proposed in \citet{Zhu2009}.  The inner disk heating causes rising ionization, which in turn initiates instabilities, including magnetorotational and thermal, which propel the gas to accrete from the disk to the star \citep[][]{Bell1994}.  The interactions of the circumstellar disk with a planet or nearby stellar partner on an eccentric orbit might be another potential triggering mechanism \citep[][]{Lodato2004, Reipurth2004, Dong2022}.  


FU Orionis type stars, which we will refer to as FUors, can exhibit a rapid increase in magnitude ($\Delta$V $>$ 5) lasting 10s of years \citep{audard2014, Fischer2023}.  
EX Lupi type stars, which we will refer to as EXors, have smaller and shorter lived outbursts.  The increased optical and near infrared magnitudes of these objects are due to an increase in the accretion rate.  The increased accretion heats a substantial part of the circumstellar disk, and in general the whole circumstellar matter, evidenced by the brightening of large-scale reflection nebulae, e.g. HBC 722 in \citet{Miller2011}.  Then, when observed at multiple wavelengths, the brightening is seen in a very broad wavelength range from the optical to the far-infrared, e.g. V1647 Ori in \citet{Muzerolle2005}, meaning a significant increase in the total bolometric luminosity as well.  In order to build an accurate profile of outbursting protostars, detailed observations with broad spectral coverage, spatial scales, and covering the epochs of before, during, and after outburst are imperative.

The question still remains on whether all young Sun-like stars undergo eruptive phases or not.  To answer this, it is important to study the properties of outbursting disks.  For instance, the disk mass ($M_{disk}$) is a particularly relevant parameter regarding whether or not the disk can be gravitationally unstable.  The $M_{disk}$ for eruptive stars have been attempted to be measured, such as in \citet{Liu2018, cieza2018,  Kospal2021}.  An ALMA survey presented in \citet{Kospal2021} found that FUor disks are more compact and more massive than disks of so-called ``regular'' (non-eruptive) Class I or Class II sources.  For reliable mass measurements, there needs to be long wavelength data/observations because of optical depth effects as disks may be optically thick even at millimeter wavelengths.  VLA, operating at cm wavelengths, is the perfect instrument for this.  The observed emission from X band data is likely the mixture of thermal and free-free emission, and observations in multiple bands and/or high spatial resolution are necessary to disentangle the different contributions.

Here, we present VLA observations of 2MASS J22352345+7517076, also known as L1251 VLA 6 and hereafter referred to as VLA 6, an embedded eruptive YSO belonging to the five-component protostellar system IRAS 22343+7501.  Embedded in the Lynds 1251 molecular cloud, the source's distance is 350$^{+46}_{-38}$ pc as presented in \citet{Kun2019}. \citet{Rosvick1995} identified a cluster of five near-infrared sources associated with IRAS 22343+7501 (RD95 A, B, C, D, and E).  The cluster is a source of several molecular outflows \citep[][]{Sato1989, Nikolic2003, Kim2015}, the Herbig–Haro jet HH 149 \citep[][]{Balzs1992}, and radio continuum jet sources (VLA 6, 7, and 10, \citealt{Reipurth2004}).  VLA 6 exists within a four component system with stars RD95A referred to as VLA 7, RD95B and RD95C referred to as VLA 10 with the young outbursting protostar referred to as VLA 6 and is identical with RD95D. We also present observations of VLA 6 that were conducted using the Faint
Object infraRed CAmera (FORCAST) with the Stratospheric Observatory for Infrared Astronomy (SOFIA) for greater insight into whether the source is declining at wavelengths of 25.3 and 31.5 $\mu$m.

\citet{onozato2015} reported the dramatic brightening of VLA 6 at mid-IR wavelengths by comparing IRAS, Akari, and WISE data.  
The history of the source's luminosity evolution is clearly outlined in \citet{Kun2019}, and especially evident in their Figure 6.  Briefly, over the course of a 35-year period, the flux tripled in the first 6 years after 1993, and then increased by a factor of 40 in the next decade until 2008.  Later, its luminosity plateaued between approximately 2008 and 2014, eventually dropping again by a factor of a few before 2016.
\citet{Kun2019} utilized an SED, corrected for interstellar extinction, and assumed the bolometric luminosity in the low-state is dominated by the central star to determine a mass range of $1.6 - 2.0 \rm ~ M_{\sun}$ for VLA 6 based on a bolometric luminosity that increased from 32 $L_\sun$ to 165 $L_\sun$ between 1983 and 2010 and its proximity to the 10$^5$-year isochrone.
The peak accretion rate, also from \citet{Kun2019}, was estimated from their accretion disk model to be slightly above $10^{-4}~\rm M_{\sun}~yr^{-1} $, typical of an FUor-type outburst.
The exact cause of the outburst, and indeed the underlying mechanism of outbursts in general, is not well constrained.

\begin{table*}
\caption{Observations}
\label{table:observations}
\centering
\begin{tabular}{lcccccc}
\hline
Date & Frequency & Array Config. & Observing Time & Baselines & No. Antennas & rms Noise\\
 & (GHz) &  & (s) & (km) & & (Jy\,beam$^{-1}$)\\
\hline
2019 Nov 08 & 33 (Ka Band) & D & 831 & $0.035 - 1.03$ & 27 & $19\times10^{-5}$\\
2020 Dec 10 & 10 (X Band) & A & 1792 & $0.68 - 36.4$ & 24 & $1\times10^{-4}$\\

\hline
\end{tabular}
\end{table*}

\begin{figure*}
  \centering
  \begin{minipage}[]{0.47\textwidth}
    \includegraphics[width=\textwidth]{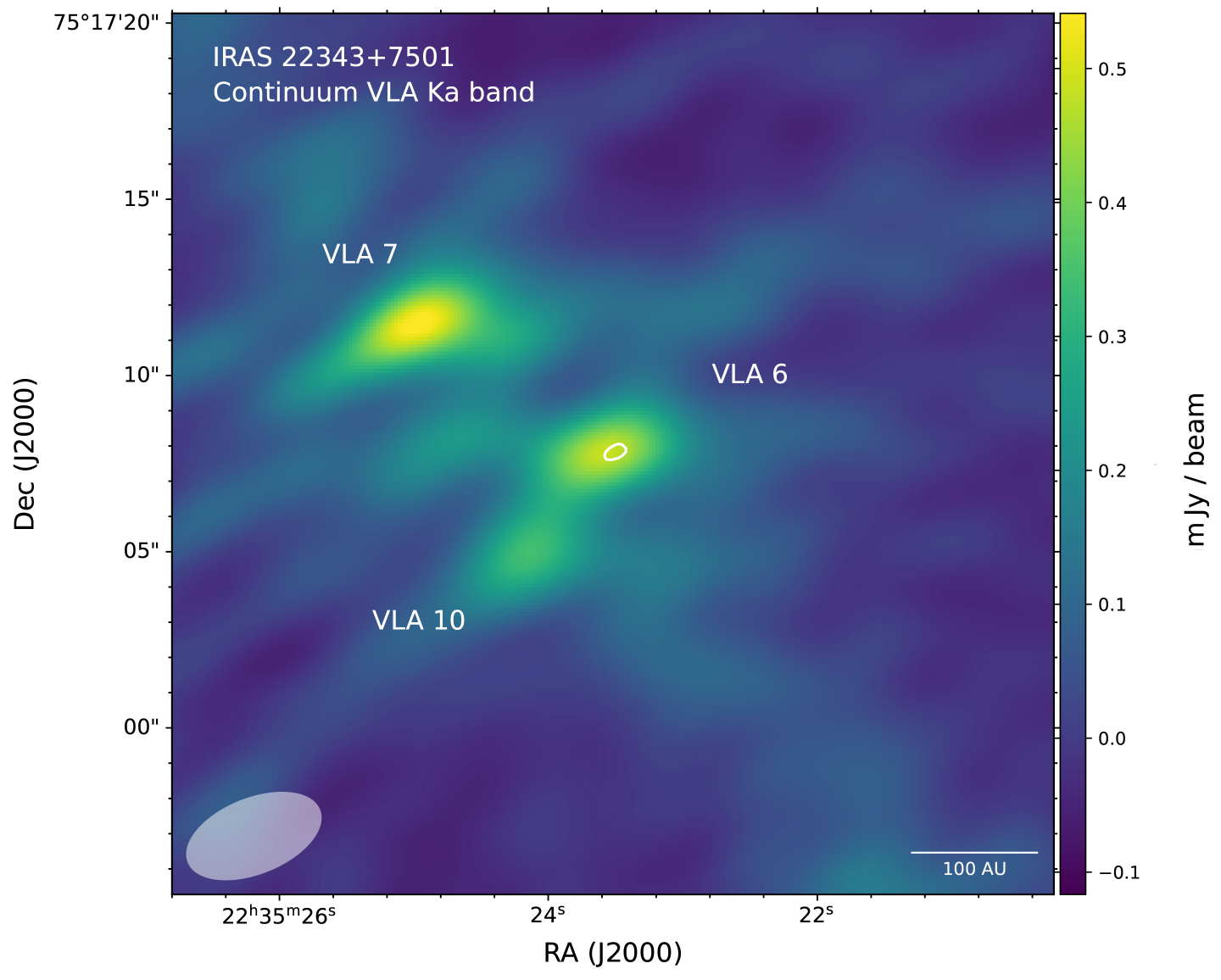}
    \caption{Continuum image of the Ka band resolved with VLA 7, VLA 10, and VLA 6 visible. White contours represent X band emission at a 4 $\sigma$ level, which clearly coincides with the source VLA 6. The white ellipse in the lower left corner represents the size and orientation of the synthesized beam for the Ka band image.}
    \label{fig:continuum_kaband}
  \end{minipage}
  \hfill
  \begin{minipage}[]{0.47\textwidth}
    \includegraphics[width=\textwidth]{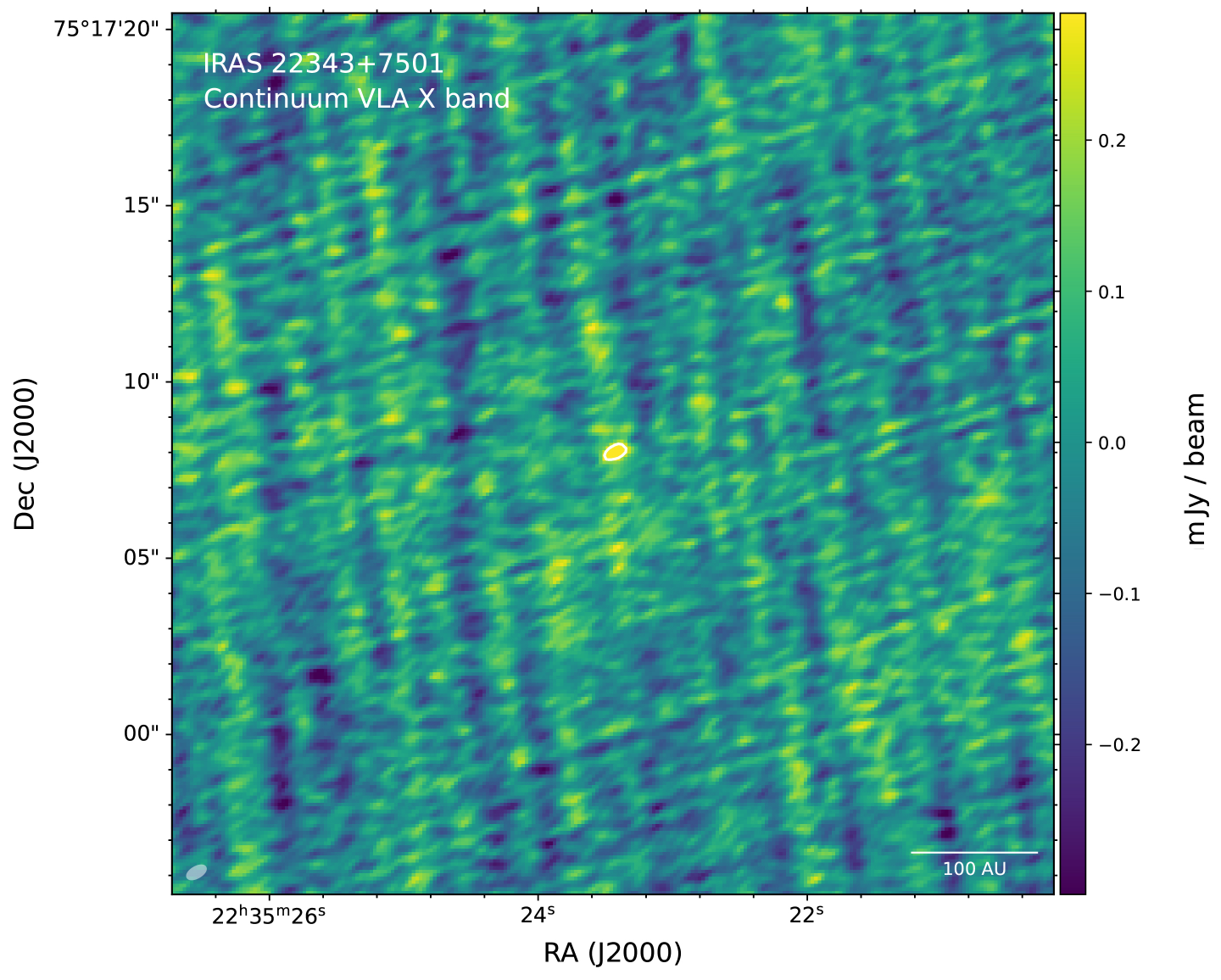}
    \caption{Continuum image of the X band marginally resolved with white contours around emission at a 4 $\sigma$ level.  The white ellipse in the lower left corner represents the size and orientation of the synthesized beam.}
    \label{fig:continuum_xband}
  \end{minipage}
\end{figure*}

The IRAS 22343+7501 cluster presents a valuable opportunity to observe a young four-component star system and the underlying effects these stars potentially have on each other, one of which is experiencing an accretion outburst.  Given the rarity of YSOs undergoing this type of accretion, VLA 6 can provide insight into the fading post-outburst process.

In this paper, we describe the new observational data in Section~\ref{sec:observations}, look at results in Section~\ref{sec:results}, examine disk parameters and constraints in Section~\ref{sec:analysis}, discuss our results in Section~\ref{sec:discussion}, and present our summary and conclusion in Section~\ref{sec:summaryconclusion}.

\section{Observations}


The IRAS 22343+7501 cluster was observed with the VLA.  We collected data at 33\,GHz data on 2019 November 08 (ID 19B-088, PI White) and at 10\,GHz on 2020 December 10 (ID 20B-096, PI White).  An overview of the observations is given in Table \ref{table:observations}.

\label{sec:observations}


\subsection{Ka Band Data}
\label{sec:kaband}


The 33\,GHz observations were centered on VLA 6 using J2000 coordinates RA = 22$^{\rm h}$ 35$^{\rm  min}$ 23.46$^{\rm  s}$ and $\delta = +75^{\circ} ~17' ~7\farcs60$.  
The data were taken using 27 antennas in the D configuration with baselines ranging from $0.035 - 1.03$ km.  The total time on-source was 831\,s. 

The observations used a Ka Band tuning setup with $4\times2.048$\,GHz basebands and rest frequency centers of 28.976\,GHz, 31.024\,GHz, 34.976\,GHz, and 37.024\,GHz.  The data yield an effective frequency of 33\,GHz (9.1 mm).  The raw data was reduced and calibrated using the {\scriptsize CASA 6.1.2-7} pipeline \citep{casa2022}, which included bandpass, flux, and phase calibrations. Quasar J0019+7327 was used for phase and bandpass calibrations and quasar 0542+498=3C147 was used as a flux calibration source.  The uncertainty of the absolute flux calibration of the VLA at these frequencies is typically $\sim 5\%$.  The pipeline calibrated data was inspected in {\scriptsize CASA} and additional flagging was performed to account for radio frequency interference (RFI).  All the significantly impacted channels and time bins were flagged and the pipeline was ran again.  
These included data from specific antennas such as ea06, ea14, ea26, and spectral windows 34-49 for antenna ea17. All of the RFI was individually removed using the {\scriptsize CASA} task \textit{flagdata}.  

The final calibrated dataset was imaged with a natural weighting and cleaned using {\scriptsize CASA}'s \textit{CLEAN} algorithm down to a threshold of $1.5\times\sigma_{RMS}$, presented in Figure~\ref{fig:continuum_kaband}.  
The observations achieve a sensitivity of $19~\rm \mu Jy~beam^{-1}$ as measured in the \textit{CLEAN}ed image.  The size of the resulting synthesized beam is $3\farcs94\times 2\farcs07$ at a position angle of $-68.3^{\circ}$.


\subsection{X Band Data}
\label{sec:xband}


The 10\,GHz observations were centered on VLA 6 using  RA = 22$^{\rm h}$ 35$^{\rm  min}$ 24.34$^{\rm  s}$ and $\delta = +75^{\circ} ~17' ~9\farcs20$. 
The data were taken using 24 antennas in the A configuration with baselines ranging from $0.68 - 36.4$\,km.  The total time on-source was 1792\,s. The observations used rest frequency centers of 8.999\,GHz and 10.999\,GHz. The data yield an effective frequency of 10\,GHz (30\,mm). 


The raw data was reduced and calibrated using the {\scriptsize CASA 6.1.2-7} pipeline \citep{casa2022}, which included bandpass, flux, and phase calibrations.  The pipeline calibrated data was inspected in {\scriptsize CASA} and additional flagging was performed to account for RFI.  All the significantly impacted channels and time bins were flagged and the pipeline was run again. The continuum image is presented in Figure~\ref{fig:continuum_xband}. The observations achieve a sensitivity of 100\,$\mu$Jy beam$^{-1}$ as measured in the \textit{CLEAN}ed image.  The size of the resulting synthesized beam is $0\farcs54\times 0\farcs26$ at a position angle of $-60.23^{\circ}$.  



\subsection{SOFIA FORCAST Data}
\label{sec:FORCASTdata}

Observations of VLA 6 were conducted using FORCAST with the SOFIA telescope on August 31, 2018 at wavelengths of 25.3 and 31.5 $\mu$m (PLANID: 06\_0165, PI Huard).  Data reduction and calibration were performed by the SOFIA team using the SOFIA data-reduction pipeline \citep{Herter2018}.

Level 4 images, which were artifact-corrected and flux-calibrated in units of Jy/pixel, were downloaded from the SOFIA archive.  Aperture photometry was performed on the images using a 12-pixel aperture and a sky annulus between 20-25 pixels.  The aperture includes the neighboring source VLA 7, and to correct for its contribution to the total flux measured in the aperture, the peak fluxes of both sources were compared and the flux ratio was used to correct the composite flux measured in the large aperture.  These observations 
can be used to further refine our understanding of the varying nature of fading at different wavelengths of VLA 6.  


\section{Results}
\label{sec:results}

Figure~\ref{fig:continuum_kaband} and Figure~\ref{fig:continuum_xband} show naturally weighted images of the combined Ka and X band data sets generated using the {\scriptsize CASA} task \textit{tclean}.  In the Ka band, three sources that correspond to sources VLA 7, VLA 10, and VLA 6 are detected, while in the X band, only VLA 6 is detected.  We used {\scriptsize CASA}'s \textit{imfit} tool to derive basic parameters such as coordinates, sizes, inclinations, position angles, and fluxes.  This enabled us to find best-fit 2D Gaussian models for each source component, which are presented in Table~\ref{table:imagingresults}.  It was not possible to determine a precise size for VLA 10 as it is measured to be nearly a point source.  The peak flux of VLA 6 is detected with the Ka band at a 120 sigma level and with the X band at a 44 sigma level.  The radii were estimated from the deconvolved best-fitted 2D Gaussian, we took the FWHM of the major axis and divided it by 2(2ln2) to obtain $\sigma$, which we use as a proxy for the radius of the disk emission at the specific wavelength based on their primary beam corrected images.  


\begin{table*}[ht]
\caption{imaging/IMFIT results}
\label{table:imagingresults}
\begin{tabular}{l ccc cc}
\toprule
 & \multicolumn{3}{c}{33\,GHz (Ka Band)} & \multicolumn{1}{c}{10\,GHz (X Band)} \\
\cmidrule(lr){2-4} \cmidrule(lr){5-6}
Parameter     &VLA 7 & VLA 10 & VLA 6 & VLA 6  &\\
\midrule
ra ('') &  22:35:24.9644   &   22:35:24.14457    & 22:35:23.5416  &  22:35:23.42834 & \\
dec ('') &   +75:17:11.4042  &  +75:17:05.04996  & +75:17:07.8022  & +75:17:08.01210 &\\
Position Angle ($^{\circ}$)  &  $121.0{\pm15}$   &  -     & $102.5{\pm8.3}$  & $109.7{\pm8.4}$\\
Peak Flux ($\mu Jy$/beam)    & $522.0{\pm19}$   & $314.2{\pm0.68}$   & $442.4{\pm3.7}$  & $617{\pm14}$ \\
Source Sizes FWHM ('')      & $2\farcs08$ x $1\farcs01$  &  - &   $1\farcs652$ x $0\farcs995$ &  $0\farcs413$ x $0\farcs282$ \\
Integrated Fluxes ($\mu Jy$) & $658{\pm40}$ & $416.2{\pm0.68}$ & $532{\pm6.7}$ & $1158{\pm380}$ \\
\bottomrule
\end{tabular}
\end{table*}

\subsection{Disk Masses}

\label{sec:diskmasses}



At millimeter wavelengths, protoplanetary disks are usually optically thin with the exception of the core regions of particularly massive systems \citep[][]{andrews2009, andrews2010}.  This means that the majority of dust grains are responsible for the observed emission, and the total flux is closely related to the total mass of small grains.  Under this assumption, the following formula can be used to determine the dust masses using millimeter fluxes:

\begin{equation} 
\label{eqn:one}
    M_{\rm dust} = \frac{d^2 F_{\nu}}{\kappa_{\nu}B_{\nu}(T)}
\end{equation}

where $d$ is the distance to the target, $F_{\nu}$ is the flux density, $\kappa_{\nu}$ is the dust opacity, and $B_{\nu}(T)$ is the Planck function evaluated at a temperature of $T$ (e.g., \citealt{andrews2013}).  
We incorporate a distance of 350$^{+46}_{-38}$\,pc based on the average parallax of the 15 known optically visible members of the L1251 star forming region included in Gaia DR2 as presented in \citet{Kun2019}, set the frequency to the 33\,GHz measurement, and make standard assumptions about the dust opacity ($\kappa_{\nu}$ = 10($\nu$/1000\,GHz)\,$cm^2 g^{-1}$; \citealt{Beckwith1990}) and beta = 1.  For the temperature, we used different values to check how its variation affects our mass estimate.  In order to obtain the total disk mass, we assume a 100:1 gas to dust ratio where we multiply our $M_{\rm dust}$ by 100.  We obtained total disk masses of 0.092, 0.055, 0.040, and 0.031  $M_\sun$ using dust temperatures of 30, 50, 70, and 90\,K, respectively.

Following from Equation~\ref{eqn:one}, we can instead look at disk mass as we vary the internal density, $\rho$, instead of $\kappa_{\nu}$.  We know that if 1/$\kappa_{\nu}$ = $\frac{4}{3}$s$\rho$, and $\Omega_{s}$ = $\pi$ $\frac{s^2}{d^2}$, then we get the following alternate equation:

%
\begin{equation}
\label{eqn:two}
    M_{\rm dust} = \frac{4}{3} \pi s^{3} \rho \frac{F_{\nu}}{B_{\nu}(T)\Omega_{s}}
\end{equation}

where $\rho$ is the internal density, $\Omega_{s}$ is the solid angle of a single grain, and $s$ is the size as presented in \citet{White2016}, where here $s$ is set to 9.08\,mm.  Similarly with Equation~\ref{eqn:one}, we obtain a disk mass by assuming a 100:1 gas-to-dust ratio where we multiply our $M_{\rm dust}$ by 100.  For grain size, it is most common to assume the size is equal to the wavelength of the observations as blackbody grains emit most efficiently at wavelengths that are about the same as their size.  
We use a range of different values for $\rho$ including 1000, 1500, 2000, 2500, and 3000\,kg/m$^{3}$.  Table~\ref{table:disk2} presents our results.



By comparing the results from Equations~\ref{eqn:one} and \ref{eqn:two}, we are able to determine how varying the density and temperature, in addition to other parameters, can affect the potential mass of the disk.  It is apparent that varying temperature and density can result in $M_{disk}$ values varying by a factor of 2. It is evident that for the Equation~\ref{eqn:one} value results, the $M_{disk}$ decreases as the temperature increases.  The same scenario similarly appears in Table~\ref{table:disk2} in addition to the $M_{disk}$ increasing as the density increases.

\citet{Kim2015} found the total circumstellar mass to be 0.033\,$M_\sun$ (adjusting for distance to that of 350\,pc assumed here) based on SMA 1.3\,mm dust continuum observations, which is at the lower end of our 95\% confidence interval.

\begin{table*}[!]
\centering
\caption{VLA 6 $M_{disk}$ calculations ($M_\sun$) varying density (kg/$m^{3}$)}
\begin{tabular}{@{}ccclll@{}}
\toprule
Temperature (K)   & $\rho$ = 1000 & $\rho$ = 1500 & $\rho$ = 2000 & $\rho$ = 2500 & $\rho$ = 3000 \\ \midrule
30 & 0.039   & 0.058   & 0.077   & 0.097   & 0.116   \\
50 & 0.023   & 0.034   & 0.046   & 0.057   & 0.069   \\
70 & 0.016   & 0.025   & 0.033   & 0.041   & 0.049   \\
90 & 0.013   & 0.019   & 0.025   & 0.032   & 0.038   \\ \bottomrule
\label{table:disk2}
\end{tabular}
\end{table*}

Now that we have been able to determine $M_{disk}$,
we are able to begin generating disk models in the \texttt{RADMC-3D} radiative transfer code, discussed in \ref{subsec:RT}, that can be compared to the images.  


\subsection{Spectral Index}


We measured the spectral index $\alpha$ of our target, VLA 6, by constructing comparisons from the two continuum spectral windows separately with $\nu_{1}$ and  $\nu_{2}$, and measuring the targets' fluxes (F) at these two frequencies.  Then, we calculated $\alpha$ using:

\begin{equation}
    \alpha = \frac{log(F_{v_1})-log(F_{v_2})}{log(v_1)-log(v_2)},
\end{equation}

and its uncertainties $\sigma$ with:

\begin{equation}
    \sigma^2 = \Big(\frac{1} {ln(v_1)-ln(v_2)} \Big)^2 \Big(\frac{ \sigma_{v_1}^2} {F_{v_1}^2} + \frac{\sigma_{v_2}^2} {F_{v_2}^2 } \Big).
\end{equation}

We calculated indices using data measured with the JCMT/SCUBA-2 in 2014, \citep{Pattle2017}, and SMA on 2007 Oct 17 \citep{Kim2015}.  The indices are also calculated for our 33\,GHz Ka band and 10\,GHz X band observations.  We found the spectral index from the 666\,GHz micron JCMT data point to 231\,GHz is  $5.670 \pm 0.018$, 231\,GHz to 33\,GHz is $2.018 \pm 0.053$, and 33\,GHz to 10\,GHz is $-0.65 \pm 0.23$.  We are able to see how the spectral index using data at 33\,GHz or above is positive, while the 33-10\,GHz slope is negative.  This could be the result of jet emissions that contaminates the flux measurement / contributes to the measured flux at lower wavelengths.  With the flux measured at 33\,GHz, and assuming that this is purely disk thermal emission, we extrapolate the flux to 10\,GHz using an adopted spectral slope (either $-3$ for blackbody or $-4.7$ for ISM-like grains).  This gives a predicted flux for 10\,GHz that can be compared to the noise level of the 10\,GHz image.  In this case, we see that the 10\,GHz flux in the VLA image is higher than what the SED predicts, which we will discuss in Section~\ref{sec:jet}.  

\begin{table*}
\centering
  \begin{threeparttable}
    \caption{SED Fluxes}
    \label{table:sed}
     \begin{tabular}{cccccc}
        \toprule
        Date of Obs. & Frequency & Flux  & eFlux & Telescope/Instrument & Ref.\\
         & (GHz) & (Jy) & (Jy) &   &   \\
        \midrule
        
        1983         & 4997 & 61.1  & 2.5   & \emph{IRAS}  &   \\
        1983         & 2998 & 77.9  & 3.1   & \emph{IRAS}  &   \\
        1993 Aug 9   & 375 & 0.710  & 0.114 & JCMT/UKT14   &  (1) \\
        1993 Aug 9   & 273 & 0.383  & 0.030 & JCMT/UKT14   &  (1)\\
        1993 Aug 9   & 231 & 0.232  & 0.024 & JCMT/UKT14   &  (1) \\

        1996 Dec 31  & 2998 & 75.5  & 5.41  & \emph{ISO/ISO}PHOT &(2) \\
        1996 Dec 31  & 2498 & 71.8  & 5.09  & \emph{ISO/ISO}PHOT &(2) \\
        1996 Dec 31  & 1499 & 105.4 & 8.48  & \emph{ISO/ISO}PHOT &(2) \\
        
        2003 Sep     & 857 & 11.80  &   -   & SHARC-II &   (8)\\
        
        2004 Sep 24  & 4283 & 57.800 & 5.744& \emph{Spitzer/}MIPS &(2)\\

        2007         & 4612 & 69.604 & 0.256& \emph{Akari/}FIS &  (3) \\
        2007         & 3331 & 84.563 & 0.042& \emph{Akari/}FIS & (3) \\
        2007         & 2141 & 118.452& 0.158& \emph{Akari/}FIS & (3)\\
        2007         & 1874 & 125.983& 0.484& \emph{Akari/}FIS &  (3)\\
        





2009 Dec 28–2010 Jan 24 &4283&96.248 & 0.739& Herschel/PACS &(4) \\


        2011 Jun     & 2998 & 82.395 & 1.309& \emph{Herschel/}PACS&(4)\\
        
        2011 Jun     & 1874 & 81.168 & 4.451& \emph{Herschel/}PACS&(4)\\
        
        2012-2014    & 666  & 14.62  &   - & JCMT/SCUBA-2 & (6)\\
        
        2012-2014    & 352  & 2.18   &   - & JCMT/SCUBA-2 & (6)\\
        
        2007 Oct 17  & 231  & 0.03523&   - & SMA & (5)\\

        2018 Aug 31  & 11850 & 75.5  & 8.9 & SOFIA/FORCAST  & (7)  \\

        2018 Aug 31  & 9517 & 90.3  & 9.8  & SOFIA/FORCAST  & (7)  \\

        2019 Nov 08  & 33   & 0.0005 &  0.0000037  & VLA & (0)\\
        
        2020 Dec 10  &  10  & 0.0012 &  0.000014  & VLA & (0)\\
        

        \bottomrule
     \end{tabular}
    \begin{tablenotes}
      \small
      \item References: 0-present work; 1-\citet{Rosvick1995}; 2-\citet{Kun2019}; 3-FIS BSC Version 2; 4-\citet{Marton2017}; 5-\citet{Kim2015}; 6-\citet{Pattle2017}; 7-\cite{Herter2018}; 8-\citet{Suresh2016}. 
    \end{tablenotes}
  \end{threeparttable}
\end{table*}

\begin{figure}[!ht]
\centering
\includegraphics[angle=0,scale=0.5]{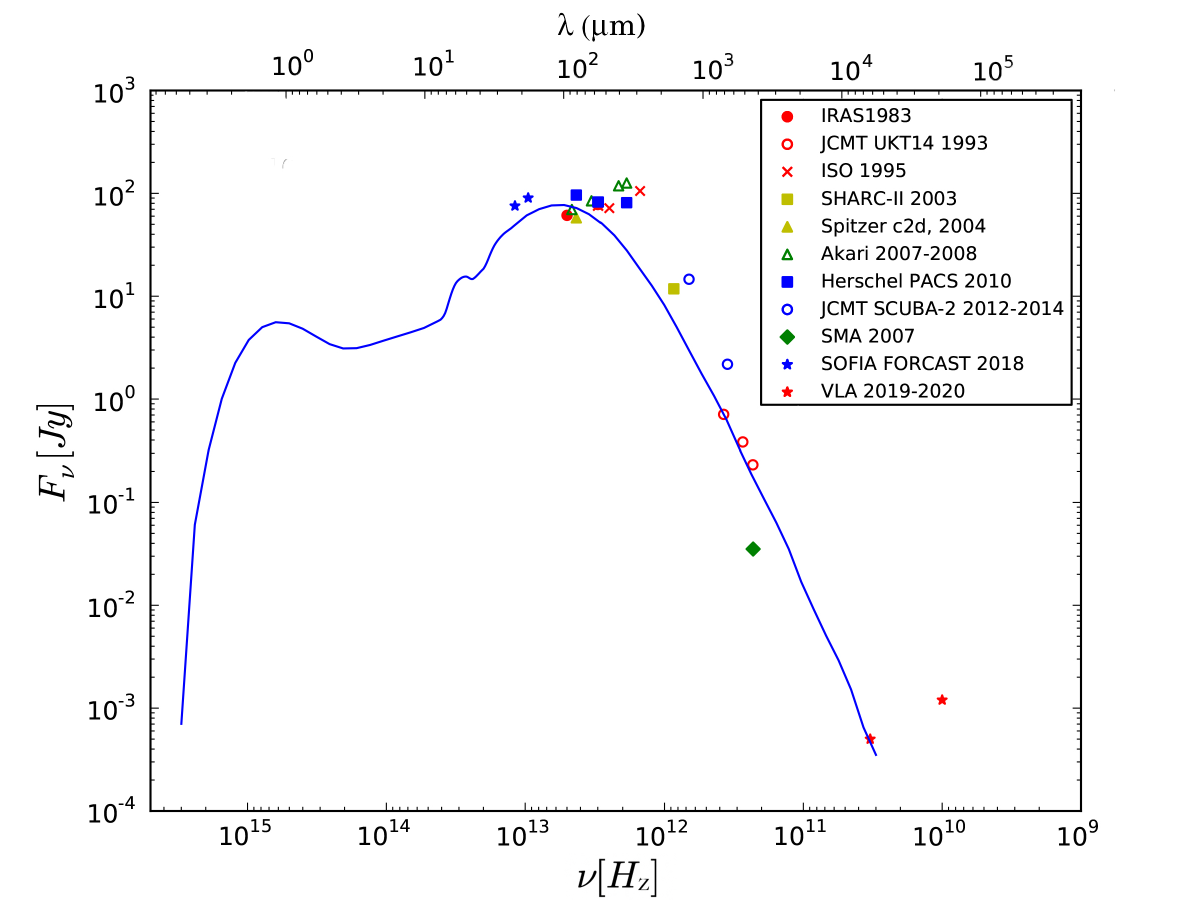}
\caption{Spectral energy distribution displaying values presented in Table~\ref{table:sed}.  The blue line represents the spectrum as performed by the \texttt{radmc3dPy} package from best-fit model. }
\label{fig:sed} 
\end{figure}

\section{Analysis}
\label{sec:analysis}

In this section, we analyze the emission of VLA 6 to characterize in detail the spatial distribution of dust in the system.  We adopt a modeling approach for the 33\,GHz data specifically taking into account the protoplanetary disk density profile equations laid out in \citet{andrews2009} and then follow the same technique in \citet{white20} that uses MCMC to constrain the most probable values.  This model is then combined with a parametric ray-tracing code to generate synthetic model images of an axisymmetric disk with an MCMC fitting algorithm, allowing us to characterize the radial distribution of dust in the system.

\subsection{Radiative Transfer Modeling}
\label{subsec:RT}

We use radiative transfer model fitting, as in \citet{Dullemond2012}, in order to constrain the disk parameters of VLA 6.  This technique gives us a more accurate model than fitting a simple Gaussian to the image shown in Figures~\ref{fig:continuum_kaband} and \ref{fig:continuum_xband}.  
The disk structure model and radiative transfer calculations are defined on a spatial grid in spherical coordinates $\{r,\phi\}$, as described in \citet{andrews2009}.  To obtain the physical parameters of the spatially resolved disk, we use the \texttt{RADMC-3D} code \citep[][]{Dullemond2012} to build radiative transfer models with the Python interface radmc3dPy\footnote{\url{https://www.ita.uni-heidelberg.de/~dullemond/software/radmc-3d/manualrmcpy/index.html}}.

The density profile of a trial protoplanetary disk model is given by:

\begin{equation}
\rho(r,\phi) =\frac{\Sigma(r,\phi)}{H_p\sqrt{(2\pi)}}\exp{\left(-\frac{z^2}{2H_p^2}\right)},
\end{equation}

where $\Sigma$ is the surface density profile, $H_p$ is the pressure scale height, and $z$ is the height above the disk midplane.  The surface density profile includes a power-law inner disk and an exponential out tapering \citep[][]{andrews2009}:

\begin{equation}
\Sigma(r) = \Sigma_c\left(\frac{r}{R_c}\right)^{-\gamma} \exp{\left\{-\left(\frac{r}{R_c} \right)^{2-\gamma}\right\}},
\end{equation}

where $R_c$ is the characteristic radius of the disk, $\gamma$ is the power-law exponent of the radial surface density profile, and $\Sigma_c$ is the surface density normalization at the inner radius.

The pressure scale height is defined as:

\begin{equation}
H_p = h_c \left( \frac{r}{R_c} \right) ^{1+\psi}, 
\end{equation}

where $h_c$ is the ratio of the pressure scale height over radius at $R_c$ and $\psi$ is the degree of flaring for the disk.  These steps follow \citet{white20}.

To generate realistic dust absorption and scattering properties, we followed the same approach as \citet{white20} and generated the \texttt{RADMC-3D} input opacity files with the {\scriptsize OpacityTool}\footnote{The OpacityTool Software was obtained from \url{https://dianaproject.wp.st-andrews.ac.uk/data-results-downloads/fortran-package/}} program \citep{toon81, woitke16}. This program calculates dust opacities by using a volume mixture of 60\% amorphous silicates \citep[e.g.,][]{dorschner95}, 15\% amorphous carbon \citep[e.g.][]{zubko96}, and a 25\% porosity.  Bruggeman mixing is used to calculate an effective refractory index, and a distribution of hollow spheres with a maximum hollow ratio of 0.8 \citep{min05} is included to avoid Mie theory scattering artifacts.  The disk is populated by $0.1 - 15000~\mu$m grains following a power-law size distribution of $s^{-3.5}$.  Finally, the total disk mass is calculated assuming a gas-to-dust mass ratio of 100:1.

\begin{figure*}
\centering
\includegraphics[angle=0,width=1.05\textwidth, scale=0.9]{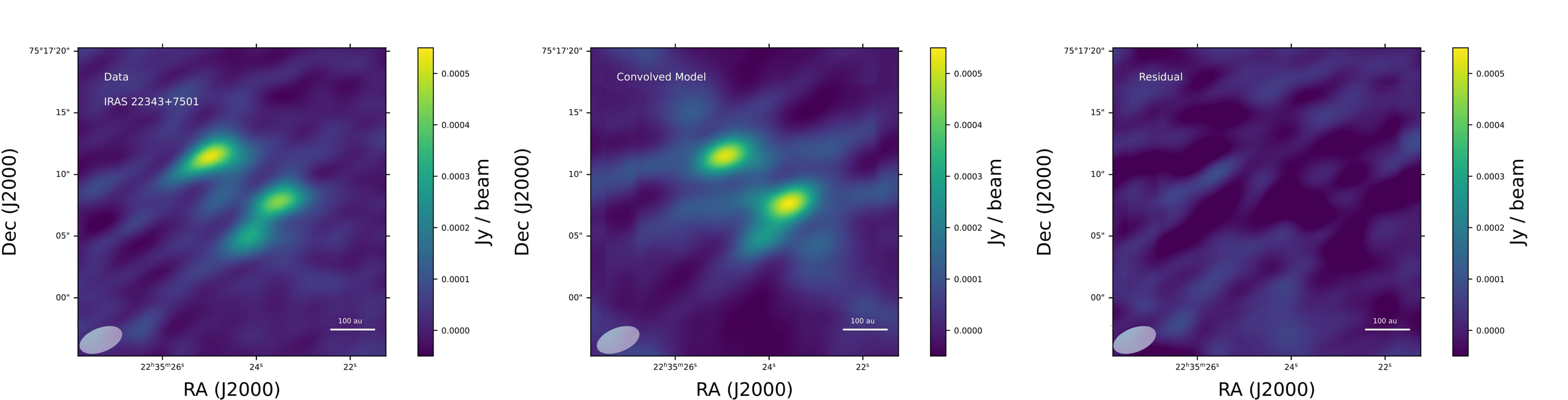}
\caption{{  (Left)} Naturally weighted Ka band VLA image of the IRAS 22343+7501 system.  { (Center)} {Convolved model image showing the most-probable model.}  { (Right)} Residual image after subtracting the convolved model from the data.}

\label{fig:dmr}
\end{figure*}

\subsection{Modeling Formalism}

To converge on the best fit model, we use Metropolis-Hastings MCMC model fitting, similar to that used in \citet{white20}.  We varied eight free parameters: stellar radius ($R_{star}$), the total disk mass ($M_{disk}$), degree of disk flaring ($\Psi$), characteristic radius ($R_{c}$), power law exponent of the surface density profile ($\gamma$), and scale height ratio ($h_{c}$). The motivation for varying the stellar radius is that we wanted to allow the model to explore potential disk properties that also consider a range of stellar radii. Setting a fixed parameter of 9.0, as hypothesized by \citet{Kun2019}, leads to parameter values that do not converge. A model is created for each individual non-point source, which for this system are VLA 6 and source VLA 7, given their locations in right ascension and declination.

Given the detailed calculation in \texttt{RADMC-3D}, a continuum image, which we will call the ``trial sky model", is then produced and projected to a trial inclination and position angle for each source.  

The model is first run on VLA 6 with a trial sky model being the result.  The trial sky model of the individual source is then combined with the sky models of other sources to create a combined sky model.  The point sources for protostars VLA 10 are included at this time, and a trial model for VLA 7 with given stand-in fixed parameters instead of free parameters is created.

The combined sky model is then attenuated by the primary beam and convolved with the synthetic beam for the observational setup in order to later compare the convolved model image with the data image.  The convolution consists of a series of Fast Fourier Transform calculations.  

To assess the likelihood of a given trial convolved model, a $\chi^2$ is calculated as

\begin{equation}
\chi^2 =  \frac{(Data - Model)^2}{\sigma^2} , 
\end{equation}

where $\sigma$ is the observed $\sigma_{rms}$ for a given observation multiplied by the synthetic beam size in pixels (see \citealt{Booth2016}).

A model is accepted if a random number drawn from a uniform distribution [0,1] is less than $\alpha$, where

\begin{equation}
\alpha =  min(e^{\frac{1}{2}(\chi_{i}^{2}-\chi_{i+1}^{2})}, 1).
\end{equation}

This process is then repeated for VLA 7 where the model runs with the free parameters to create a new trial sky model.  The trial sky model of this individual source is then combined with the sky models of other sources.  The point sources for protostars VLA 10 are included at this time, and the trial model for VLA 6 with the previously determined best-fit parameters instead of free parameters is created.  


For both VLA 6 and VLA 7, we ran 100 chains with 1,000 links each beyond the burn-in period of 100 links.  The most probable values are summarized in Table~\ref{table:mcmc}.  The 95\% confidence regions are also presented, indicated from the posterior distributions in Figure~\ref{fig:corner_D}.  These confidence intervals provide uncertainties for the MCMC fit and correspond roughly to $2\sigma$ uncertainties.  Figure~\ref{fig:dmr} shows the data image (left) compared with the synthesized best-fit model (center right) and the residuals (right).  The lack of residuals after the best-fit model is subtracted from the data verifies a good fit {\it a posteriori}.

\begin{deluxetable}{lcccccc}
\centering
\tablewidth{0pt}  
\tablecaption{MCMC Fitting Results 
\label{table:mcmc}
}
\tablehead{
    \hline
    \multicolumn{1}{l}{Parameter} & 
    \multicolumn{2}{c}{VLA 6} &
    \multicolumn{2}{c}{VLA 7} & \\
    \cline{2-3} \cline{4-5} \cline{6-7}
    &Best Fit & 95\% regions &
    Best Fit & 95\% regions\\
 }
\startdata
$M_{disk}$ ($M_{\odot}$) & 0.070   & [0.031, 0.2]  & 0.103  & [0.052,0.301]  \\
$R_{c}$ (au)           & 260.0   & [110, 340]    & 365.8  & [200, 440]    \\ 
$\gamma$               & -0.110  & [-1.78, 1.1]  & -0.992 & [-1.9, 0.99]  \\ 
$h_{c}$ (au)           & 0.130   & [0.014, 0.08] & 0.103  & [0.04, 0.08] \\ 
$R_{star}$ ($R_{\odot}$)      & 4.5     & [0.81, 7.9]   & 4$^a$  &  –  \\ 
$\Psi$                 & 0.250   & [0.011, 0.48] & 0.204  & [0.11, 0.5] \\
\enddata
\tablecomments{$^a$ $R_{star}$ is a fixed parameter for VLA 7, so there are no 95\% confidence regions. }
\end{deluxetable}

\section{Discussion}
\label{sec:discussion}

We found that the resulting total disk mass of $0.070 \rm ~ M_{\sun}$ from the MCMC fitting in comparison to the dust calculations in Section~\ref{sec:diskmasses} is accurately reflected.  The disk mass value is similar to the resulting value when temperature is set to 50\,K in Equation~\ref{eqn:one}.  The value is also similarly consistent when temperature is 50\,K and density is 3,000\,$kg/m^3$ in Equation~\ref{eqn:two}.  This value also closely resembles to when the temperature is 30\,K and density is 2,000\,$kg/m^3$.

\subsection{Classification}

\begin{figure*}
\centering
\includegraphics[angle=0,scale=0.65]{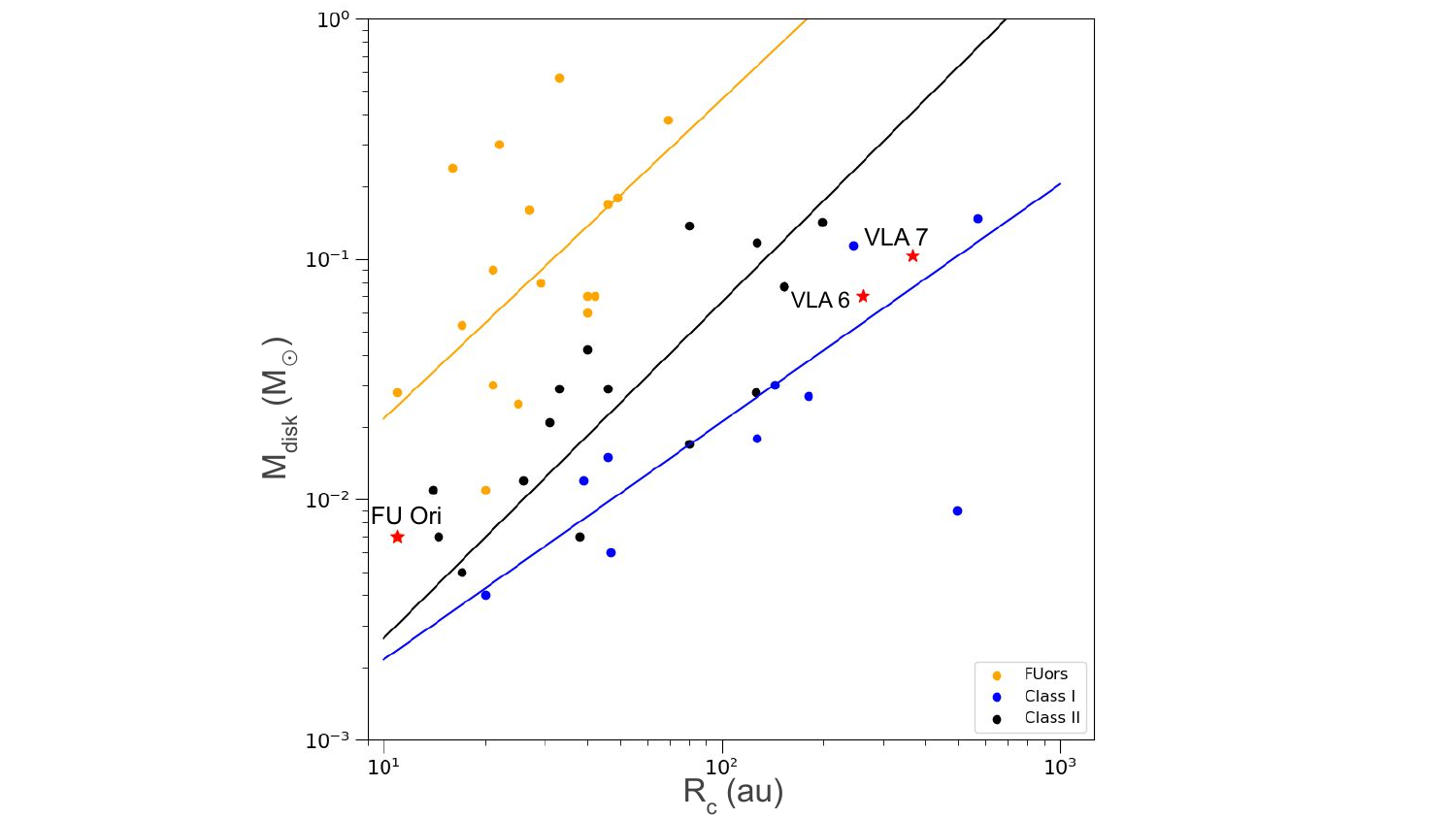}
\caption{Total $M_{disk}$ (in $M_{\sun}$) as a function of characteristic radii.  The orange points are the FUors presented in \citet{perez2010, perez2020, cieza2018, Kospal2021}.  The black points are T Tauri disks from \citet{andrews2010}.  The blue points are Class I disks from \citet{Sheehan2014, sheehan2017}.  FU Ori, VLA 6, and VLA 7 are labeled.  The linear fits are presented through the solid lines, where $M_{disk} \propto R^{1.33}_{c}$ represents the FUors, $M_{disk} \propto R^{0.99}_{c}$ represents Class I and $M_{disk} \propto R^{1.40}_{c}$ represents Class II.}
\label{fig:mass_radius}
\end{figure*}

In Figure~\ref{fig:mass_radius}, we show disk masses as a function of characteristic disk radii for FUors and non-eruptive Class I and Class II sources, reproduced from \citet{Kospal2021}.
The results of our MCMC fit show that the most probable values for the mass and radius are more consistent with Class I as seen in Figure~\ref{fig:mass_radius}.  In comparison with VLA 7, the eruptive star has less massive, smaller disk with flatter surface density profile, higher scale height, and more flaring based on the parameters in Table~\ref{table:mcmc}.  However, the confidence intervals largely overlap, implying no significant difference between these two disks.


The results presented in \citet{Kospal2021} suggest that FUor disks differ from Class I/II disks in the distribution of their masses and radii.  
The FUors presented have a median FUor disk radius of 27\,au and median FUor $M_{disk}$ of 0.08\,$M_{\sun}$.  The Class I disks that are plotted in blue and were sampled from \citet{Sheehan2014,sheehan2017} have a median disk radius of 127\,au and median $M_{disk}$ of 0.02\,$M_{\sun}$.  Class II comparisons were drawn from \citet{andrews2010}, and show a median disk radius of 40\,au and median $M_{disk}$ of 0.03\,$M_{\sun}$.

From these trends, it is evident that FUor disks tend to be more massive and compact than regular Class I/II disks.  However, we find that with a characteristic disk radius of 260 au, VLA 6 is larger for its disk mass of 0.07\,$M_{\sun}$, coinciding more with regular Class I disks, rather than FUor disks.  We also find that the mass-radius ratio of VLA 6 and VLA 7 are similar.  It is interesting to note that despite these similarities, VLA 7 is not an outbursting source like VLA 6.  Thus, it appears that the mass-radius ratio is not a property that distinguishes FUor-like protostars from non-outbursting YSOs in all cases.  This result is in contrast with \citet{Kospal2021} that raised the possibility that FUor eruptions were specific to a Class of YSO's with a unique compact disk structure that undergoes a different evolutionary path.  The larger radius of VLA 6, instead, suggests that there may be a population of outbursting stars that deviate from these assumed FUor properties.  We propose that these deviations may be dependent on the cause of outburst (as discussed in Section~\ref{subsec:causes}).

All FUor and Class II sources presented in Figure~\ref{fig:mass_radius} were modeled similarly with \texttt{RADMC-3D} and their parameters were constrained with MCMC fitting as our own modeling for VLA 6.  The data in \citet{Kospal2021} used 1.3\,mm observations while we present here 9\,mm data, therefore, there will be a potential difference in the obtained disk mass or disk radius.  The Class I disks were also modeled with \texttt{RADMC-3D} though did not constrain the parameters with the same approach, using a power law surface density profile without exponential taper.  Though the modeling formalism differs from that of the FUor samples, \citet{Kospal2021} found that setting the characteristic radius as the outer disk radius, $R_{disk}$ from \citet{sheehan2017}, produced similar spectral profiles and nearly identical total disk masses.  Given the lack of studies on Class I disks with the same model as presented in this paper, these coinciding results make the parameters from \citet{sheehan2017} most comparable to that of the FUor samples.



Historically, young outbursting protostars were classified by well defined categories.  Observed diversity, as described in \citet{Fischer2023}, shows that this is no longer necessarily the case.  Therefore, it is possible there are many young outbursting stars that are similar to VLA 6 but are yet to be discovered and labeled as such.  According to the classification scheme set by \citet{ConnelleyReipurth2018}, VLA 6 is classified as a ``peculiar'' object that was similar but distinct from classical FUors; in the following discussion we explore several additional characteristics of VLA 6 in the context of other FUors for reference, yet the exact designation should not impede the interpretation of this particular source.


\subsection{Gravitational Instability / Causes of Outburst}
\label{subsec:causes}

\begin{figure*}[!]
\centering
\includegraphics[angle=0,scale=1.0]{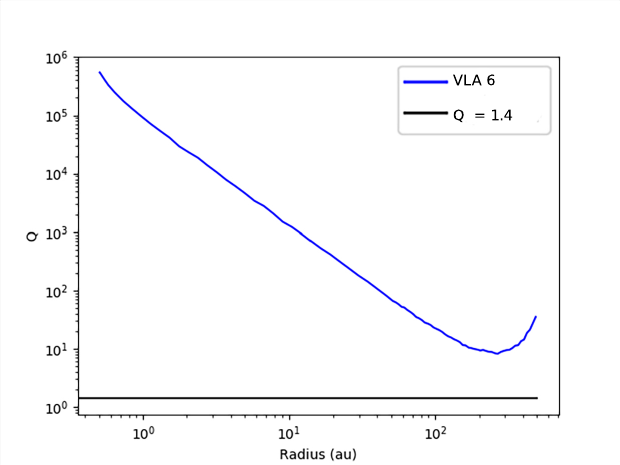}
\caption{GI plot with the upper limit of Toomre's $Q$ parameter of 1.4 for VLA 6 represented by the black line.  The calculations for VLA 6 are represented by the blue line.  }
\label{fig:gi}
\end{figure*}

The larger and small-scale structure of FUor disks can reveal the source of the episodic accretion that causes the outbursts.  While the exact mechanisms that drive increased disk material accretion onto the protostar are unknown, it could be a result of the disk becoming gravitationally unstable, possibly in conjunction with a magnetorotational instability.
GI is one possible mechanism for outbursts \citep[][]{Vorobyov2006, Zhu2009, Kuffmeier2018}.  \citet{Kospal2021} checked the criterion for GI in a sample of FUor disks and found that about 2/3 of them may be gravitationally unstable.  There are multiple metrics to calculate the GI, the more detailed of which is Toomre's $Q$ parameter.  GI occurs if the value of Toomre's $Q$ parameter is $Q = c_s \Omega / \pi G \Sigma\leq1.4$, where $c_s$ is the sound speed, $\Omega$ is the epicyclic frequency for a Keplerian disk, and $\Sigma$ is the column density at a given radius in the disk.  Based on our best-fitting radiative transfer model, we calculated the Toomre Q parameter and the results are presented in Figure~\ref{fig:gi}.  

We found that VLA 6 is gravitationally stable, consistent with the other 1/3 of the FUor disks that were examined in \citet{Kospal2021}.  This contrasts with predictions in \citet{Kun2019} that suggested GI was the cause of a $10^{-3} M_{\sun}$ clump accretion onto the star within 10\,au.  Appearances of GI earlier on in the history of the outbursting process may be considered.  Present results of there being no evidence of GI implies that there are other underlying causes for the enhanced accretion.

According to \citet{hartmann1996}, mass is added onto the central star so rapidly that the magnetosphere which may launch an X wind is likely crushed out of existence, implying that the strong outflows that still occur during these outbursts must be caused by an extended disk wind. Several molecular outflows originate from the cluster \citep{Sato1989, Nikolic2003, Kim2015}; \citet{Balazs2004} reports on the Herbig-Haro jet HH 149, and radio continuum jet sources have been observed in VLA 6, 7, and 10 by \citet{Reipurth2004}.


Close encounters in multiple systems and flybys have been shown to trigger outbursts \citep[][]{Borchert2022}.  As predicted by models, interactions between the stars and their disks for close binaries cause accretion pulses to occur \citep[][]{Artymowicz1996}.  Recent findings indicate that for less massive disks ($M_{disk}\leq 0.1 M_{\sun}$), close stellar encounters and thermal instability are more likely causes of outbursts, whereas disk fragmentation, GI, and magnetorotational instability are more mass-dependent mechanisms \citep[][]{cieza2018}.  The low disk mass of 0.07\,M$_{\odot}$  may explain why the mass-radius ratio of VLA 6 deviates from that of the FUors sampled in \citet{Kospal2021} as seen in Figure~\ref{fig:mass_radius}.  Further investigation with higher angular resolution will have to be conducted in order to determine the cause of the outburst in this system.

\subsection{Disk Flaring}
\label{sec:diskflaring}
\begin{figure}
\centering
\includegraphics[angle=0,scale=.6]{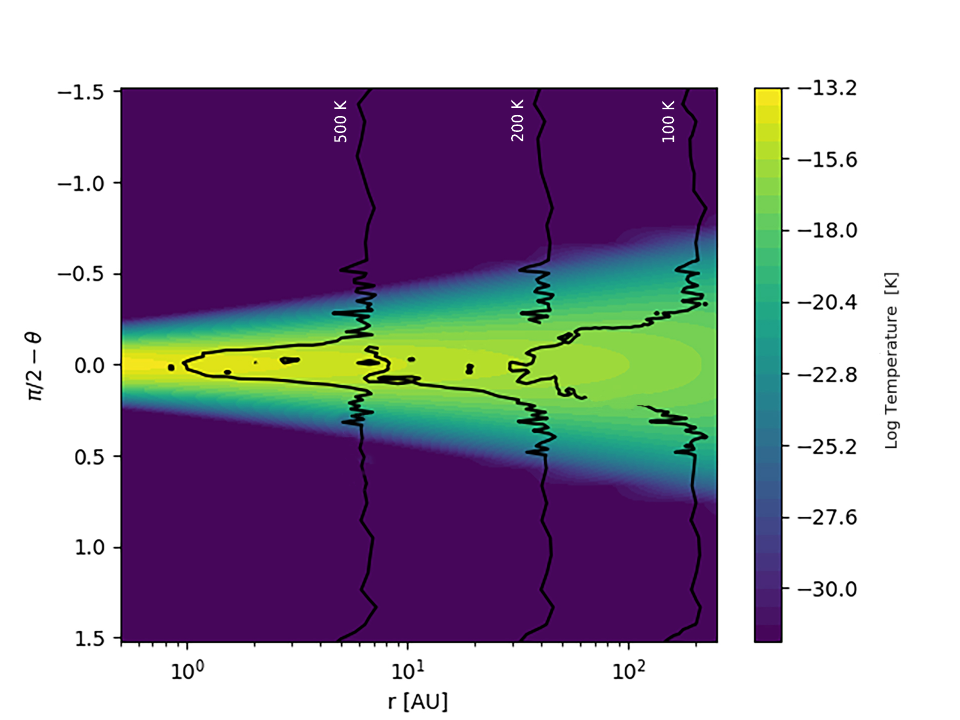}
\caption{2D slice of the model disk's density structure for VLA 6. }
\label{fig:slice}
\end{figure}

Figure~\ref{fig:slice} shows a 2D slice of the model disk's density structure for VLA 6 as a function of radius.  The disk is flared due to the outburst heating up the disk which increases the vertical mixing and therefore making the disk thicker.  The black contour lines correspond to the disk temperature.  These move outward from the outburst as the central effective temperature increases.


\subsection{Possible Jet Emission} 
\label{sec:jet}

We can confidently determine that the X band detection coincides with the outbursting VLA 6 source, looking at the contours in Figure~\ref{fig:continuum_kaband}.  The X band data can give insight into emission observed from episodic jets. At this frequency, free-free emission is most likely from an ionized jet \citep[][]{Madrid2015}.  Other possibilities for emission include non-thermal emission that may come from the disk, disk wind, or stellar wind.  Given the distance, 1\,mJy in the X band may be too bright to be dust emission as the dust mass is too high.  The X band flux is higher than what the SED modeling in Fig~\ref{fig:sed} would imply at that frequency, and it is more likely to consider that free-free emission is obscured by dust at high-frequency.  We use our best-fit radiative transfer model to produce the spectral energy distribution presented in Figure~\ref{fig:sed} (blue line) from our MCMC results.  The flux value predicted by this model for a disk with $0.070 \rm ~ M_{\sun}$ mass would produce a thermal emission of 60\,$\mu$Jy at 10\,GHz, whereas what we observe is about 1200\,$\mu$Jy, producing a factor that can go up to 60 higher.  

Based on our radiative transfer modeling of the disk emission, the thermal disk flux is negligible in the 8-12\,GHz range.  At 10\,GHz, the model predicts a thermal flux of 60\,$\mu$Jy, but we measure 1158\,$\mu$Jy, so the thermal flux is probably only about 5\% of the measured total flux at 10\,GHz.  Therefore, the whole range of free–free emission is closely constrained by the measured X band flux density at any given moment \citep[][]{Liu2018}.  The flux densities at higher frequencies that cannot be explained by free-free emission must be assigned to dust thermal emission, as these are well fitted by our radiative transfer model in Figure~\ref{fig:sed}.  

Previously, free-free emission from ionized gas was used to fit the VLA data for FU Ori \citep[][]{Liu2019, Liu2021} non-thermal components related to radiatively or thermally ionized gas close to the outbursting star.  

For the quiescent disks, such bright free-free emission is most likely related to ionized jets. In accretion bursts, the disk mid-plane can be thermally ionized.  A post-bust source may have a much lower luminosity than in the bursting phase, but the temperature should not be so different (e.g., in \citealt{Kun2019}; the luminosity is only a factor of ~5 different, meaning that the temperature may be only different by 1.5 times). If there is some thermally ionized gas in the bursting phase, this may exist directly after the post-burst phase.


\subsection{Fading}
\label{sec:fading}

There is speculation as to whether the source is in decline or not.  On one hand, the WISE data showed a decline between the years 2015 and 2017 as seen in \citet{Kun2019}, but then another increase in brightness since then.  The SOFIA FORCAST data from 2018 are consistent with the SED constructed with data from Herschel PACS in 2010 and Spitzer IRAC in 2009-2010, among others.  Therefore, it seems that the source brightness varies depending on the wavelength being observed, and additionally the fading may happen differently at different wavelengths.  

\section{Summary and Conclusion}
\label{sec:summaryconclusion}

We present 33\,GHz and 10\,GHz (9\,mm and 3\,cm) radio interferometric data of the IRAS 22343+7501 protostellar multiple system, which contains the outbursting Class I protostar VLA 6.  We firmly detected VLA 6 at 33\,GHz and spatially resolve its emission, with have a significant detection of 10\,GHz but did not spatially resolve it.  We adopt a modeling approach that combines a parametric ray-tracing code to generate synthetic model images of an axisymmetric disk with an MCMC fitting algorithm, allowing us to characterize the radial distribution of dust in the system.
To converge on the best fit model, we use Metropolis-Hastings MCMC model fitting. We were able to constrain parameters for the source using MCMC, finding that the resulting disk mass is $0.070 \rm ~ M_{\sun}$, while the characteristic disk radius is 260\,au.  This value is similar to dust mass calculations assuming optically thin emission.

The results of our MCMC fit show that the most probable values for the mass and radius of VLA 6 are more consistent with (regular) Class I sources, rather than Class II.  This confirmation is important to our understanding of the properties of the fading outburst system.

\acknowledgments
We express appreciation to Hauyu Baobab Liu for advice about the X band emission, and Sebasti\'{a}n P\'{e}rez for discussion about the data methods.  A.N. acknowledges support from the National Science Foundation Graduate Research Fellowship under Grant No. 2234683. Any opinion, findings, and conclusions or recommendations expressed in this material are those of the authors(s) and do not necessarily reflect the views of the National Science Foundation. A.N. was supported by the Research Experiences for Undergraduates program of the National Science Foundation.  The National Radio Astronomy Observatory is a facility of the National Science Foundation operated under cooperative agreement by Associated Universities, Inc. A.N. acknowledges support from the NASA CT Space Grant Consortium. This project has received funding from the European Research Council (ERC) under the European Union's Horizon 2020 research and innovation programme under grant agreement No 716155 (SACCRED).  J.A.W. contributed to this manuscript in his personal capacity.  The contents of this manuscript do not reflect the views of EarthDaily Analytics, its subsidiaries, or employees.  This paper makes use of the following VLA data:  VLA/19B-088 and VLA/20B-096.  The National Radio Astronomy Observatory is a facility of the National Science Foundation operated under cooperative agreement by Associated Universities, Inc.  Based [in part] on observations made with the NASA/DLR Stratospheric Observatory for Infrared Astronomy (SOFIA).  SOFIA is jointly operated by the Universities Space Research Association, Inc. (USRA), under NASA contract NNA17BF53C, and the Deutsches SOFIA Institut (DSI) under DLR contract 50 OK 2002 to the University of Stuttgart.

\textit{Software:} \texttt{APLpy} \citep{Robitaille2012, Robitaille2019}, \texttt{Astropy} \citep{astropy2013},  \texttt{CASA} \citep{mcmullin2007}, \texttt{Matplotlib} \citep{hunter2007},  \texttt{NumPy} \citep{vanderWalt2011},  \texttt{Pandas} \citep{mckinney2010}, \texttt{RADMC-3D} \citep{Dullemond2012}, \texttt{radmc3dPy}.

\clearpage

\appendix
\label{appendix}

\begin{figure}[t]
\centering
\includegraphics[angle=0,scale=0.4]{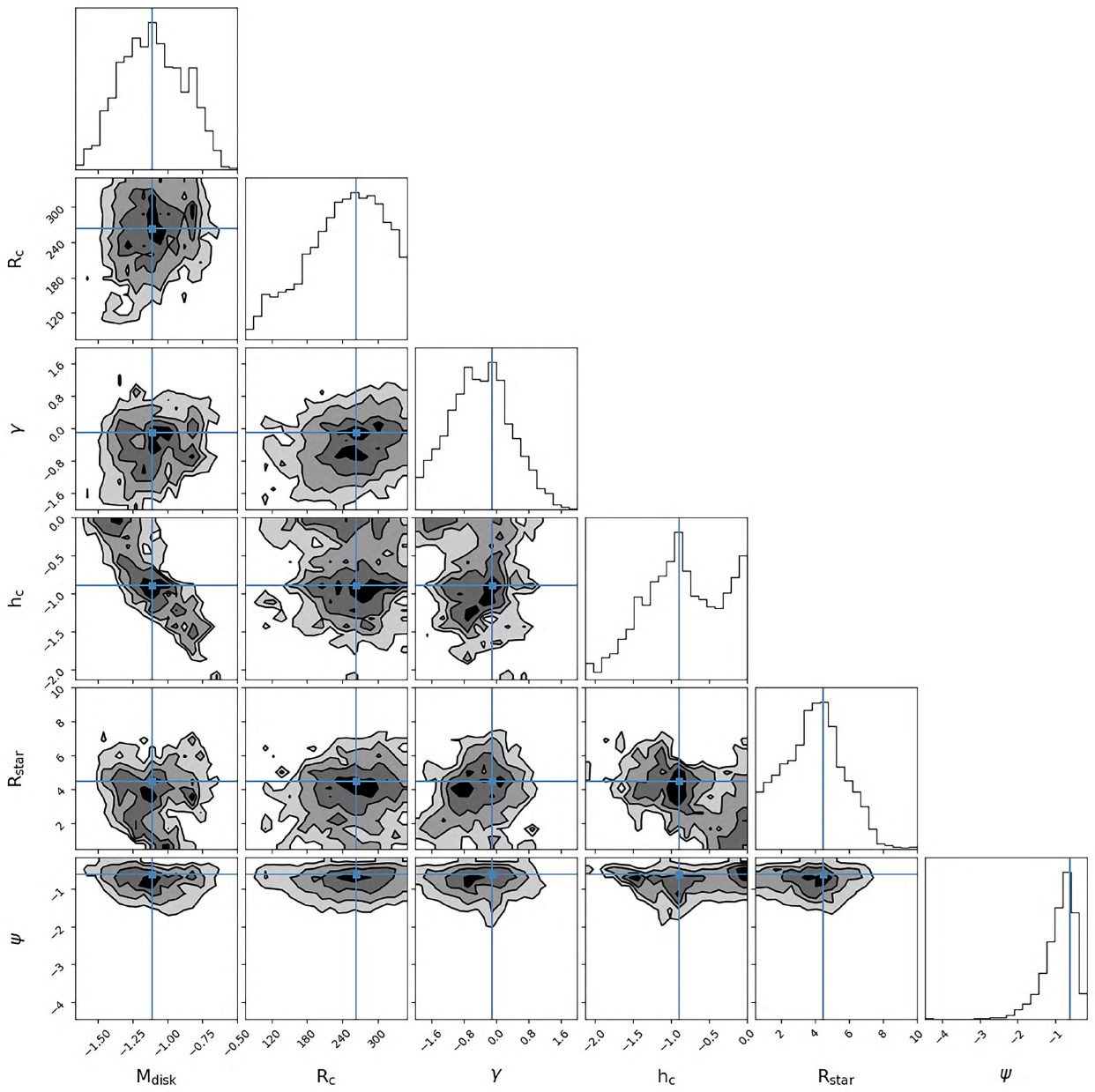}
\caption{Posterior distributions for the MCMC chains for VLA 6 after removing burn-in.  Histograms for each parameter are plotted at the top of the corresponding column, while the plots for the rest of the grid show the distribution of walkers across slices through parameter space for the corresponding pair of parameters.  The best-fit value for each parameter is plotted with a blue line. }
\label{fig:corner_D}
\end{figure}

\begin{figure}[t]
\centering
\includegraphics[angle=0,scale=0.3]{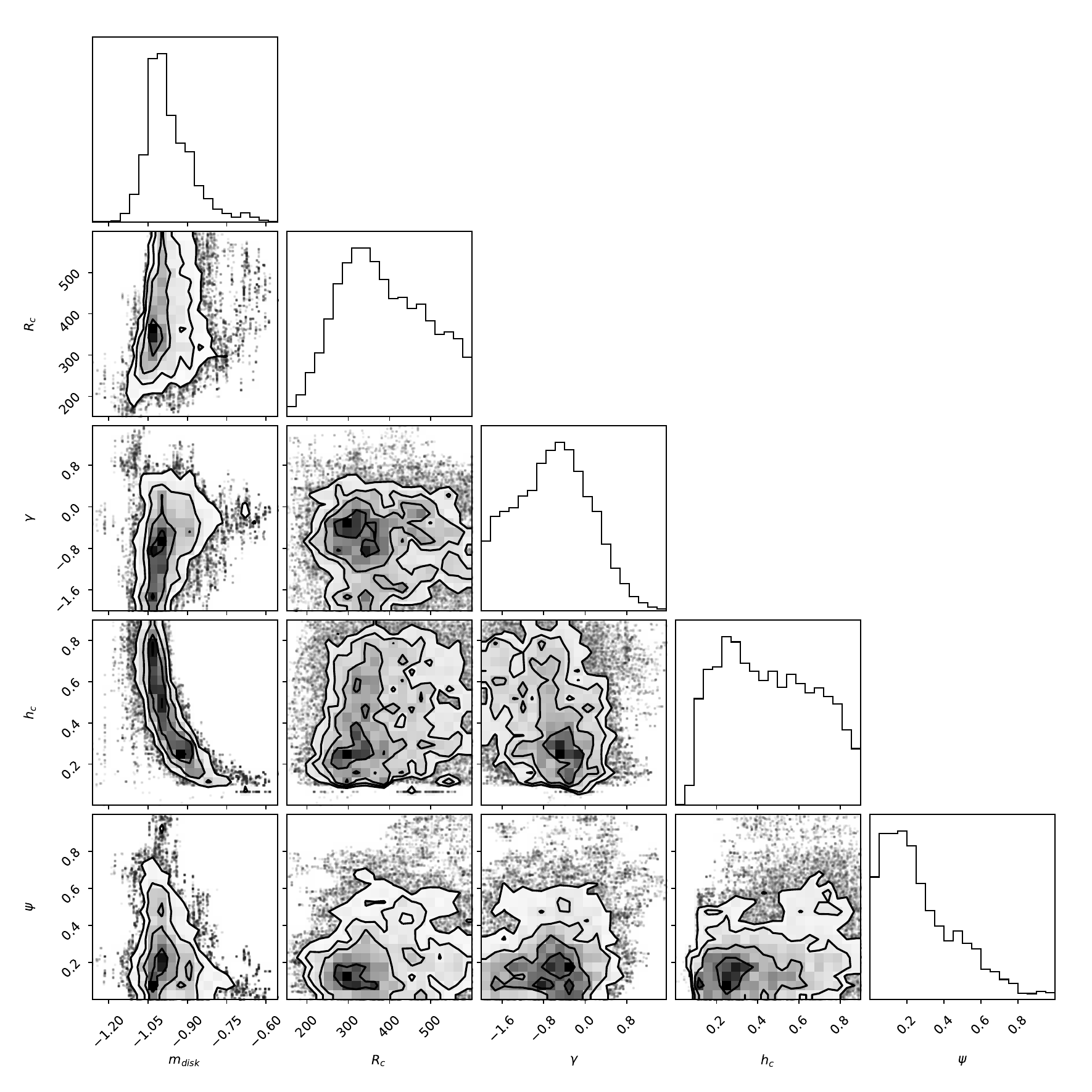}
\caption{Same as Figure~\ref{fig:corner_D} but for VLA 7.}
\label{fig:corner_A}
\end{figure}

In this Appendix, we show the Posterior distributions for the MCMC chains for VLA 6 (Figure~\ref{fig:corner_D}) and VLA 7 (Figure~\ref{fig:corner_A}).

\bibliographystyle{apj}{}
\bibliography{L1251}

\begin{thebibliography}{}
\expandafter\ifx\csname natexlab\endcsname\relax\def\natexlab#1{#1}\fi

\bibitem[{{{\'A}brah{\'a}m} {et~al.}(2009){{\'A}brah{\'a}m}, {Juh{\'a}sz}, {Dullemond}, {K{\'o}sp{\'a}l}, {van Boekel}, {Bouwman}, {Henning}, {Mo{\'o}r}, {Mosoni}, {Sicilia-Aguilar}, \& {Sipos}}]{abraham2009}
{{\'A}brah{\'a}m}, P., {Juh{\'a}sz}, A., {Dullemond}, C.~P., {et~al.} 2009, \nat, 459, 224

\bibitem[{{Andrews} {et~al.}(2013){Andrews}, {Rosenfeld}, {Kraus}, \& {Wilner}}]{andrews2013}
{Andrews}, S.~M., {Rosenfeld}, K.~A., {Kraus}, A.~L., \& {Wilner}, D.~J. 2013, \apj, 771, 129

\bibitem[{{Andrews} {et~al.}(2009){Andrews}, {Wilner}, {Hughes}, {Qi}, \& {Dullemond}}]{andrews2009}
{Andrews}, S.~M., {Wilner}, D.~J., {Hughes}, A.~M., {Qi}, C., \& {Dullemond}, C.~P. 2009, \apj, 700, 1502

\bibitem[{{Andrews} {et~al.}(2010){Andrews}, {Wilner}, {Hughes}, {Qi}, \& {Dullemond}}]{andrews2010}
---. 2010, \apj, 723, 1241

\bibitem[{{Arce} {et~al.}(2007){Arce}, {Shepherd}, {Gueth}, {Lee}, {Bachiller}, {Rosen}, \& {Beuther}}]{Arce2007}
{Arce}, H.~G., {Shepherd}, D., {Gueth}, F., {et~al.} 2007, in Protostars and Planets V, ed. B.~{Reipurth}, D.~{Jewitt}, \& K.~{Keil}, 245

\bibitem[{{Artymowicz} \& {Lubow}(1996)}]{Artymowicz1996}
{Artymowicz}, P., \& {Lubow}, S.~H. 1996, \apjl, 467, L77

\bibitem[{{Astropy Collaboration} {et~al.}(2018){Astropy Collaboration}, {Price-Whelan}, {Sip{\H{o}}cz}, {G{\"u}nther}, {Lim}, {Crawford}, {Conseil}, {Shupe}, {Craig}, {Dencheva}, {Ginsburg}, {Vand erPlas}, {Bradley}, {P{\'e}rez-Su{\'a}rez}, {de Val-Borro}, {Aldcroft}, {Cruz}, {Robitaille}, {Tollerud}, {Ardelean}, {Babej}, {Bach}, {Bachetti}, {Bakanov}, {Bamford}, {Barentsen}, {Barmby}, {Baumbach}, {Berry}, {Biscani}, {Boquien}, {Bostroem}, {Bouma}, {Brammer}, {Bray}, {Breytenbach}, {Buddelmeijer}, {Burke}, {Calderone}, {Cano Rodr{\'\i}guez}, {Cara}, {Cardoso}, {Cheedella}, {Copin}, {Corrales}, {Crichton}, {D'Avella}, {Deil}, {Depagne}, {Dietrich}, {Donath}, {Droettboom}, {Earl}, {Erben}, {Fabbro}, {Ferreira}, {Finethy}, {Fox}, {Garrison}, {Gibbons}, {Goldstein}, {Gommers}, {Greco}, {Greenfield}, {Groener}, {Grollier}, {Hagen}, {Hirst}, {Homeier}, {Horton}, {Hosseinzadeh}, {Hu}, {Hunkeler}, {Ivezi{\'c}}, {Jain}, {Jenness}, {Kanarek}, {Kendrew}, {Kern}, {Kerzendorf}, {Khvalko}, {King}, {Kirkby}, {Kulkarni},
  {Kumar}, {Lee}, {Lenz}, {Littlefair}, {Ma}, {Macleod}, {Mastropietro}, {McCully}, {Montagnac}, {Morris}, {Mueller}, {Mumford}, {Muna}, {Murphy}, {Nelson}, {Nguyen}, {Ninan}, {N{\"o}the}, {Ogaz}, {Oh}, {Parejko}, {Parley}, {Pascual}, {Patil}, {Patil}, {Plunkett}, {Prochaska}, {Rastogi}, {Reddy Janga}, {Sabater}, {Sakurikar}, {Seifert}, {Sherbert}, {Sherwood-Taylor}, {Shih}, {Sick}, {Silbiger}, {Singanamalla}, {Singer}, {Sladen}, {Sooley}, {Sornarajah}, {Streicher}, {Teuben}, {Thomas}, {Tremblay}, {Turner}, {Terr{\'o}n}, {van Kerkwijk}, {de la Vega}, {Watkins}, {Weaver}, {Whitmore}, {Woillez}, {Zabalza}, \& {Astropy Contributors}}]{astropy2013}
{Astropy Collaboration}, {Price-Whelan}, A.~M., {Sip{\H{o}}cz}, B.~M., {et~al.} 2018, \aj, 156, 123

\bibitem[{{Audard} {et~al.}(2014){Audard}, {{\'A}brah{\'a}m}, {Dunham}, {Green}, {Grosso}, {Hamaguchi}, {Kastner}, {K{\'o}sp{\'a}l}, {Lodato}, {Romanova}, {Skinner}, {Vorobyov}, \& {Zhu}}]{audard2014}
{Audard}, M., {{\'A}brah{\'a}m}, P., {Dunham}, M.~M., {et~al.} 2014, in Protostars and Planets VI, ed. H.~{Beuther}, R.~S. {Klessen}, C.~P. {Dullemond}, \& T.~{Henning}, 387

\bibitem[{{Bae} {et~al.}(2014){Bae}, {Hartmann}, {Zhu}, \& {Nelson}}]{Bae2014}
{Bae}, J., {Hartmann}, L., {Zhu}, Z., \& {Nelson}, R.~P. 2014, \apj, 795, 61

\bibitem[{{Bal{\'a}zs} {et~al.}(2004){Bal{\'a}zs}, {{\'A}brah{\'a}m}, {Kun}, {Kelemen}, \& {T{\'o}th}}]{Balazs2004}
{Bal{\'a}zs}, L.~G., {{\'A}brah{\'a}m}, P., {Kun}, M., {Kelemen}, J., \& {T{\'o}th}, L.~V. 2004, \aap, 425, 133

\bibitem[{{Balazs} {et~al.}(1992){Balazs}, {Eisloeffel}, {Holl}, {Kelemen}, \& {Kun}}]{Balzs1992}
{Balazs}, L.~G., {Eisloeffel}, J., {Holl}, A., {Kelemen}, J., \& {Kun}, M. 1992, \aap, 255, 281

\bibitem[{{Beckwith} {et~al.}(1990){Beckwith}, {Sargent}, {Chini}, \& {Guesten}}]{Beckwith1990}
{Beckwith}, S. V.~W., {Sargent}, A.~I., {Chini}, R.~S., \& {Guesten}, R. 1990, \aj, 99, 924

\bibitem[{{Bell} \& {Lin}(1994)}]{Bell1994}
{Bell}, K.~R., \& {Lin}, D.~N.~C. 1994, \apj, 427, 987

\bibitem[{{Bell} {et~al.}(1995){Bell}, {Lin}, {Hartmann}, \& {Kenyon}}]{Bell1995}
{Bell}, K.~R., {Lin}, D.~N.~C., {Hartmann}, L.~W., \& {Kenyon}, S.~J. 1995, \apj, 444, 376

\bibitem[{{Booth} {et~al.}(2016){Booth}, {Jord{\'a}n}, {Casassus}, {Hales}, {Dent}, {Faramaz}, {Matr{\`a}}, {Barkats}, {Brahm}, \& {Cuadra}}]{Booth2016}
{Booth}, M., {Jord{\'a}n}, A., {Casassus}, S., {et~al.} 2016, \mnras, 460, L10

\bibitem[{{Borchert} {et~al.}(2022){Borchert}, {Price}, {Pinte}, \& {Cuello}}]{Borchert2022}
{Borchert}, E. M.~A., {Price}, D.~J., {Pinte}, C., \& {Cuello}, N. 2022, \mnras, 510, L37

\bibitem[{{CASA Team} {et~al.}(2022){CASA Team}, {Bean}, {Bhatnagar}, {Castro}, {Donovan Meyer}, {Emonts}, {Garcia}, {Garwood}, {Golap}, {Villalba}, {Harris}, {Hayashi}, {Hoskins}, {Hsieh}, {Jagannathan}, {Kawasaki}, {Keimpema}, {Kettenis}, {Lopez}, {Marvil}, {Masters}, {McNichols}, {Mehringer}, {Miel}, {Moellenbrock}, {Montesino}, {Nakazato}, {Ott}, {Petry}, {Pokorny}, {Raba}, {Rau}, {Schiebel}, {Schweighart}, {Sekhar}, {Shimada}, {Small}, {Steeb}, {Sugimoto}, {Suoranta}, {Tsutsumi}, {van Bemmel}, {Verkouter}, {Wells}, {Xiong}, {Szomoru}, {Griffith}, {Glendenning}, \& {Kern}}]{casa2022}
{CASA Team}, {Bean}, B., {Bhatnagar}, S., {et~al.} 2022, \pasp, 134, 114501

\bibitem[{Cieza {et~al.}(2018)Cieza, Ru{\'\i}z-Rodr{\'\i}guez, Perez, Casassus, Williams, Zurlo, Principe, Hales, Prieto, Tobin, {et~al.}}]{cieza2018}
Cieza, L.~A., Ru{\'\i}z-Rodr{\'\i}guez, D., Perez, S., {et~al.} 2018, Monthly Notices of the Royal Astronomical Society, 474, 4347

\bibitem[{{Connelley} \& {Reipurth}(2018)}]{ConnelleyReipurth2018}
{Connelley}, M.~S., \& {Reipurth}, B. 2018, \apj, 861, 145

\bibitem[{{Cruz-S{\'a}enz de Miera} {et~al.}(2022){Cruz-S{\'a}enz de Miera}, {K{\'o}sp{\'a}l}, {{\'A}brah{\'a}m}, {Park}, {Nagy}, {Siwak}, {Kun}, {Fiorellino}, {Szab{\'o}}, {Antoniucci}, {Giannini}, {Nisini}, {Szabados}, {Kriskovics}, {Ordasi}, {Szak{\'a}ts}, {Vida}, {Vink{\'o}}, {Zieli{\'n}ski}, {Wyrzykowski}, {Garc{\'\i}a-{\'A}lvarez}, {Dr{\'o}{\.z}d{\.z}}, {Og{\l}oza}, \& {Sonbas}}]{Miera2022}
{Cruz-S{\'a}enz de Miera}, F., {K{\'o}sp{\'a}l}, {\'A}., {{\'A}brah{\'a}m}, P., {et~al.} 2022, \apj, 927, 125

\bibitem[{{Dong} {et~al.}(2022){Dong}, {Liu}, {Cuello}, {Pinte}, {{\'A}brah{\'a}m}, {Vorobyov}, {Hashimoto}, {K{\'o}sp{\'a}l}, {Chiang}, {Takami}, {Chen}, {Dunham}, {Fukagawa}, {Green}, {Hasegawa}, {Henning}, {Pavlyuchenkov}, {Pyo}, \& {Tamura}}]{Dong2022}
{Dong}, R., {Liu}, H.~B., {Cuello}, N., {et~al.} 2022, Nature Astronomy, 6, 331

\bibitem[{Dorschner {et~al.}(1995)Dorschner, Begemann, Henning, Jaeger, \& Mutschke}]{dorschner95}
Dorschner, J., Begemann, B., Henning, T., Jaeger, C., \& Mutschke, H. 1995, Astronomy and Astrophysics, 300, 503

\bibitem[{{Dullemond} {et~al.}(2012){Dullemond}, {Juhasz}, {Pohl}, {Sereshti}, {Shetty}, {Peters}, {Commercon}, \& {Flock}}]{Dullemond2012}
{Dullemond}, C.~P., {Juhasz}, A., {Pohl}, A., {et~al.} 2012, {RADMC-3D: A multi-purpose radiative transfer tool}, ascl:1202.015

\bibitem[{{Fischer} {et~al.}(2023){Fischer}, {Hillenbrand}, {Herczeg}, {Johnstone}, {Kospal}, \& {Dunham}}]{Fischer2023}
{Fischer}, W.~J., {Hillenbrand}, L.~A., {Herczeg}, G.~J., {et~al.} 2023, in Astronomical Society of the Pacific Conference Series, Vol. 534, Astronomical Society of the Pacific Conference Series, ed. S.~{Inutsuka}, Y.~{Aikawa}, T.~{Muto}, K.~{Tomida}, \& M.~{Tamura}, 355

\bibitem[{{Galv{\'a}n-Madrid} {et~al.}(2015){Galv{\'a}n-Madrid}, {Rodr{\'\i}guez}, {Liu}, {Costigan}, {Palau}, {Zapata}, \& {Loinard}}]{Madrid2015}
{Galv{\'a}n-Madrid}, R., {Rodr{\'\i}guez}, L.~F., {Liu}, H.~B., {et~al.} 2015, \apjl, 806, L32

\bibitem[{{Hartmann} \& {Kenyon}(1996)}]{hartmann1996}
{Hartmann}, L., \& {Kenyon}, S.~J. 1996, \araa, 34, 207

\bibitem[{{Herbig}(1989)}]{Herbig1989}
{Herbig}, G.~H. 1989, in European Southern Observatory Conference and Workshop Proceedings, Vol.~33, European Southern Observatory Conference and Workshop Proceedings, 233--246

\bibitem[{{Herter} {et~al.}(2018){Herter}, {Adams}, {Gull}, {Schoenwald}, {Keller}, {Pirger}, {Henderson}, {Stacey}, {Nikola}, {De Buizer}, {Vacca}, \& {Ennico}}]{Herter2018}
{Herter}, T.~L., {Adams}, J.~D., {Gull}, G.~E., {et~al.} 2018, Journal of Astronomical Instrumentation, 7, 1840005

\bibitem[{{Hillenbrand} {et~al.}(2019){Hillenbrand}, {Reipurth}, {Connelley}, {Cutri}, \& {Isaacson}}]{Hillenbrand2019}
{Hillenbrand}, L.~A., {Reipurth}, B., {Connelley}, M., {Cutri}, R.~M., \& {Isaacson}, H. 2019, \aj, 158, 240

\bibitem[{{Hillenbrand} {et~al.}(2018){Hillenbrand}, {Contreras Pe{\~n}a}, {Morrell}, {Naylor}, {Kuhn}, {Cutri}, {Rebull}, {Hodgkin}, {Froebrich}, \& {Mainzer}}]{Hillenbrand2018}
{Hillenbrand}, L.~A., {Contreras Pe{\~n}a}, C., {Morrell}, S., {et~al.} 2018, \apj, 869, 146

\bibitem[{{Hodapp} {et~al.}(2019){Hodapp}, {Reipurth}, {Pettersson}, {Tonry}, {Denneau}, {Vallely}, {Shappee}, {Armstrong}, {Connelley}, {Kochanek}, {Fausnaugh}, {Chini}, {Haas}, \& {Sobrino Figaredo}}]{Hodapp2019}
{Hodapp}, K.~W., {Reipurth}, B., {Pettersson}, B., {et~al.} 2019, \aj, 158, 241

\bibitem[{{Hodapp} {et~al.}(2020){Hodapp}, {Denneau}, {Tucker}, {Shappee}, {Huber}, {Payne}, {Do}, {Lin}, {Connelley}, {Varricatt}, {Tonry}, {Chambers}, \& {Magnier}}]{Hodapp2020}
{Hodapp}, K.~W., {Denneau}, L., {Tucker}, M., {et~al.} 2020, \aj, 160, 164

\bibitem[{{Hunter}(2007)}]{hunter2007}
{Hunter}, J.~D. 2007, Computing in Science and Engineering, 9, 90

\bibitem[{{Kim} {et~al.}(2015){Kim}, {Lee}, {Choi}, {Bourke}, {Evans}, {Di Francesco}, {Cieza}, {Dunham}, \& {Kang}}]{Kim2015}
{Kim}, J., {Lee}, J.-E., {Choi}, M., {et~al.} 2015, \apjs, 218, 5

\bibitem[{{K{\'o}sp{\'a}l} {et~al.}(2021){K{\'o}sp{\'a}l}, {Cruz-S{\'a}enz de Miera}, {White}, {{\'A}brah{\'a}m}, {Chen}, {Csengeri}, {Dong}, {Dunham}, {Feh{\'e}r}, {Green}, {Hashimoto}, {Henning}, {Hogerheijde}, {Kudo}, {Liu}, {Takami}, \& {Vorobyov}}]{Kospal2021}
{K{\'o}sp{\'a}l}, {\'A}., {Cruz-S{\'a}enz de Miera}, F., {White}, J.~A., {et~al.} 2021, arXiv e-prints, arXiv:2106.14409

\bibitem[{{Kuffmeier} {et~al.}(2018){Kuffmeier}, {Frimann}, {Jensen}, \& {Haugb{\o}lle}}]{Kuffmeier2018}
{Kuffmeier}, M., {Frimann}, S., {Jensen}, S.~S., \& {Haugb{\o}lle}, T. 2018, \mnras, 475, 2642

\bibitem[{{Kun} {et~al.}(2019){Kun}, {{\'A}brah{\'a}m}, {Acosta Pulido}, {Mo{\'o}r}, \& {Prusti}}]{Kun2019}
{Kun}, M., {{\'A}brah{\'a}m}, P., {Acosta Pulido}, J.~A., {Mo{\'o}r}, A., \& {Prusti}, T. 2019, \mnras, 483, 4424

\bibitem[{{Liu} {et~al.}(2018){Liu}, {Dunham}, {Pascucci}, {Bourke}, {Hirano}, {Longmore}, {Andrews}, {Carrasco-Gonz{\'a}lez}, {Forbrich}, {Galv{\'a}n-Madrid}, {Girart}, {Green}, {Ju{\'a}rez}, {K{\'o}sp{\'a}l}, {Manara}, {Palau}, {Takami}, {Testi}, \& {Vorobyov}}]{Liu2018}
{Liu}, H.~B., {Dunham}, M.~M., {Pascucci}, I., {et~al.} 2018, \aap, 612, A54

\bibitem[{{Liu} {et~al.}(2019){Liu}, {M{\'e}rand}, {Green}, {P{\'e}rez}, {Hales}, {Yang}, {Dunham}, {Hasegawa}, {Henning}, {Galv{\'a}n-Madrid}, {K{\'o}sp{\'a}l}, {Takami}, {Vorobyov}, \& {Zhu}}]{Liu2019}
{Liu}, H.~B., {M{\'e}rand}, A., {Green}, J.~D., {et~al.} 2019, \apj, 884, 97

\bibitem[{{Liu} {et~al.}(2021){Liu}, {Tsai}, {Chen}, {Liu}, {Zhang}, {Ma}, {Elbakyan}, {Green}, {Hales}, {Liu}, {Takami}, {P{\'e}rez}, {Vorobyov}, \& {Yang}}]{Liu2021}
{Liu}, H.~B., {Tsai}, A.-L., {Chen}, W.~P., {et~al.} 2021, \apj, 923, 270

\bibitem[{{Lodato} \& {Clarke}(2004)}]{Lodato2004}
{Lodato}, G., \& {Clarke}, C.~J. 2004, \mnras, 353, 841

\bibitem[{{Marton} {et~al.}(2017){Marton}, {Calzoletti}, {Perez Garcia}, {Kiss}, {Paladini}, {Altieri}, {Sanchez Portal}, {Kidger}, \& {the Herschel Point Source Catalogue Working Group}}]{Marton2017}
{Marton}, G., {Calzoletti}, L., {Perez Garcia}, A.~M., {et~al.} 2017, arXiv e-prints, arXiv:1705.05693

\bibitem[{{McMullin} {et~al.}(2007){McMullin}, {Waters}, {Schiebel}, {Young}, \& {Golap}}]{mcmullin2007}
{McMullin}, J.~P., {Waters}, B., {Schiebel}, D., {Young}, W., \& {Golap}, K. 2007, in Astronomical Society of the Pacific Conference Series, Vol. 376, Astronomical Data Analysis Software and Systems XVI, ed. R.~A. {Shaw}, F.~{Hill}, \& D.~J. {Bell}, 127

\bibitem[{{Miller} {et~al.}(2011){Miller}, {Hillenbrand}, {Covey}, {Poznanski}, {Silverman}, {Kleiser}, {Rojas-Ayala}, {Muirhead}, {Cenko}, {Bloom}, {Kasliwal}, {Filippenko}, {Law}, {Ofek}, {Dekany}, {Rahmer}, {Hale}, {Smith}, {Quimby}, {Nugent}, {Jacobsen}, {Zolkower}, {Velur}, {Walters}, {Henning}, {Bui}, {McKenna}, {Kulkarni}, {Klein}, {Kandrashoff}, \& {Morton}}]{Miller2011}
{Miller}, A.~A., {Hillenbrand}, L.~A., {Covey}, K.~R., {et~al.} 2011, \apj, 730, 80

\bibitem[{Min {et~al.}(2005)Min, Hovenier, \& de~Koter}]{min05}
Min, M., Hovenier, J., \& de~Koter, A. 2005, Astronomy \& Astrophysics, 432, 909

\bibitem[{{Muzerolle} {et~al.}(2005){Muzerolle}, {Megeath}, {Flaherty}, {Gordon}, {Rieke}, {Young}, \& {Lada}}]{Muzerolle2005}
{Muzerolle}, J., {Megeath}, S.~T., {Flaherty}, K.~M., {et~al.} 2005, \apjl, 620, L107

\bibitem[{{Nagy} {et~al.}(2023){Nagy}, {Park}, {{\'A}brah{\'a}m}, {K{\'o}sp{\'a}l}, {Cruz-S{\'a}enz de Miera}, {Kun}, {Siwak}, {Szab{\'o}}, {Szil{\'a}gyi}, {Fiorellino}, {Giannini}, {Lee}, {Lee}, {Marton}, {Szabados}, {Vitali}, {Andrzejewski}, {Gromadzki}, {Hodgkin}, {Jab{\l}o{\'n}ska}, {Mendez}, {Merc}, {Michniewicz}, {Miko{\l}ajczyk}, {Pylypenko}, {Ratajczak}, {Wyrzykowski}, {Zejmo}, \& {Zieli{\'n}ski}}]{Nagy2023}
{Nagy}, Z., {Park}, S., {{\'A}brah{\'a}m}, P., {et~al.} 2023, \mnras, 524, 3344

\bibitem[{{Nikoli{\'c}} {et~al.}(2003){Nikoli{\'c}}, {Johansson}, \& {Harju}}]{Nikolic2003}
{Nikoli{\'c}}, S., {Johansson}, L.~E.~B., \& {Harju}, J. 2003, \aap, 409, 941

\bibitem[{{Onozato} {et~al.}(2015){Onozato}, {Ita}, {Ono}, {Fukagawa}, {Yanagisawa}, {Izumiura}, {Nakada}, \& {Matsunaga}}]{onozato2015}
{Onozato}, H., {Ita}, Y., {Ono}, K., {et~al.} 2015, \pasj, 67, 39

\bibitem[{{Pattle} {et~al.}(2017){Pattle}, {Ward-Thompson}, {Kirk}, {Di Francesco}, {Kirk}, {Mottram}, {Keown}, {Buckle}, {Beaulieu}, {Berry}, {Broekhoven-Fiene}, {Currie}, {Fich}, {Hatchell}, {Jenness}, {Johnstone}, {Nutter}, {Pineda}, {Quinn}, {Salji}, {Tisi}, {Walker-Smith}, {Hogerheijde}, {Bastien}, {Bresnahan}, {Butner}, {Chen}, {Chrysostomou}, {Coud{\'e}}, {Davis}, {Drabek-Maunder}, {Duarte-Cabral}, {Fiege}, {Friberg}, {Friesen}, {Fuller}, {Graves}, {Greaves}, {Gregson}, {Holland}, {Joncas}, {Knee}, {Mairs}, {Marsh}, {Matthews}, {Moriarty-Schieven}, {Mowat}, {Rawlings}, {Richer}, {Robertson}, {Rosolowsky}, {Rumble}, {Sadavoy}, {Thomas}, {Tothill}, {Viti}, {White}, {Wouterloot}, {Yates}, \& {Zhu}}]{Pattle2017}
{Pattle}, K., {Ward-Thompson}, D., {Kirk}, J.~M., {et~al.} 2017, \mnras, 464, 4255

\bibitem[{{Pech} {et~al.}(2010){Pech}, {Loinard}, {Chandler}, {Rodr{\'\i}guez}, {D'Alessio}, {Brogan}, {Wilner}, \& {Ho}}]{Pech2010}
{Pech}, G., {Loinard}, L., {Chandler}, C.~J., {et~al.} 2010, \apj, 712, 1403

\bibitem[{P{\'e}rez {et~al.}(2010)P{\'e}rez, Lamb, Woody, Carpenter, Zauderer, Isella, Bock, Bolatto, Carlstrom, Culverhouse, {et~al.}}]{perez2010}
P{\'e}rez, L.~M., Lamb, J.~W., Woody, D.~P., {et~al.} 2010, The Astrophysical Journal, 724, 493

\bibitem[{{P{\'e}rez} {et~al.}(2020){P{\'e}rez}, {Hales}, {Liu}, {Zhu}, {Casassus}, {Williams}, {Zurlo}, {Cuello}, {Cieza}, \& {Principe}}]{perez2020}
{P{\'e}rez}, S., {Hales}, A., {Liu}, H.~B., {et~al.} 2020, \apj, 889, 59

\bibitem[{{Rab} {et~al.}(2017){Rab}, {G{\"u}del}, {Padovani}, {Kamp}, {Thi}, {Woitke}, \& {Aresu}}]{rab2017}
{Rab}, C., {G{\"u}del}, M., {Padovani}, M., {et~al.} 2017, \aap, 603, A96

\bibitem[{{Reipurth} {et~al.}(2004){Reipurth}, {Rodr{\'\i}guez}, {Anglada}, \& {Bally}}]{Reipurth2004}
{Reipurth}, B., {Rodr{\'\i}guez}, L.~F., {Anglada}, G., \& {Bally}, J. 2004, \aj, 127, 1736

\bibitem[{{Robitaille}(2019)}]{Robitaille2019}
{Robitaille}, T. 2019, {APLpy v2.0: The Astronomical Plotting Library in Python}, Zenodo, doi:10.5281/zenodo.2567476

\bibitem[{{Robitaille} \& {Bressert}(2012)}]{Robitaille2012}
{Robitaille}, T., \& {Bressert}, E. 2012, {APLpy: Astronomical Plotting Library in Python}, Astrophysics Source Code Library, record ascl:1208.017, ascl:1208.017

\bibitem[{{Rosvick} \& {Davidge}(1995)}]{Rosvick1995}
{Rosvick}, J.~M., \& {Davidge}, T.~J. 1995, \pasp, 107, 49

\bibitem[{{Sato} \& {Fukui}(1989)}]{Sato1989}
{Sato}, F., \& {Fukui}, Y. 1989, \apj, 343, 773

\bibitem[{{Semkov} {et~al.}(2021){Semkov}, {Ibryamov}, \& {Peneva}}]{Semkov2021}
{Semkov}, E., {Ibryamov}, S., \& {Peneva}, S. 2021, Symmetry, 13, 2433

\bibitem[{{Sheehan} \& {Eisner}(2014)}]{Sheehan2014}
{Sheehan}, P.~D., \& {Eisner}, J.~A. 2014, \apj, 791, 19

\bibitem[{{Sheehan} \& {Eisner}(2017)}]{sheehan2017}
---. 2017, \apj, 851, 45

\bibitem[{{Siwak} {et~al.}(2023){Siwak}, {Hillenbrand}, {K{\'o}sp{\'a}l}, {{\'A}brah{\'a}m}, {Giannini}, {De}, {Mo{\'o}r}, {Szil{\'a}gyi}, {Jan{\'\i}k}, {Koen}, {Park}, {Nagy}, {Cruz-S{\'a}enz de Miera}, {Fiorellino}, {Marton}, {Kun}, {Lucas}, {Udalski}, \& {Szab{\'o}}}]{Siwak2023}
{Siwak}, M., {Hillenbrand}, L.~A., {K{\'o}sp{\'a}l}, {\'A}., {et~al.} 2023, \mnras, 524, 5548

\bibitem[{{Suresh} {et~al.}(2016){Suresh}, {Dunham}, {Arce}, {Evans}, {Bourke}, {Merello}, \& {Wu}}]{Suresh2016}
{Suresh}, A., {Dunham}, M.~M., {Arce}, H.~G., {et~al.} 2016, \aj, 152, 36

\bibitem[{{Szegedi-Elek} {et~al.}(2020){Szegedi-Elek}, {{\'A}brah{\'a}m}, {Wyrzykowski}, {Kun}, {K{\'o}sp{\'a}l}, {Chen}, {Marton}, {Mo{\'o}r}, {Kiss}, {P{\'a}l}, {Szabados}, {Varga}, {Varga-Vereb{\'e}lyi}, {Andreas}, {Bachelet}, {Bischoff}, {B{\'o}di}, {Breedt}, {Burgaz}, {Butterley}, {Carrasco}, {{\v{C}}epas}, {Damljanovic}, {Gezer}, {Godunova}, {Gromadzki}, {Gurgul}, {Hardy}, {Hildebrandt}, {Hoffmann}, {Hundertmark}, {Ihanec}, {Janulis}, {Kalup}, {Kaczmarek}, {K{\"o}nyves-T{\'o}th}, {Krezinger}, {Kruszy{\'n}ska}, {Littlefair}, {Maskoli{\={u}}nas}, {M{\'e}sz{\'a}ros}, {Miko{\l}ajczyk}, {Mugrauer}, {Netzel}, {Ordasi}, {Pak{\v{s}}tien{\.{e}}}, {Rybicki}, {S{\'a}rneczky}, {Seli}, {Simon}, {{\v{S}}i{\v{s}}kauskait{\.{e}}}, {S{\'o}dor}, {Sokolovsky}, {Stenglein}, {Street}, {Szak{\'a}ts}, {Tomasella}, {Tsapras}, {Vida}, {Zdanavi{\v{c}}ius}, {Zieli{\'n}ski}, {Zieli{\'n}ski}, \& {Zi{\'o}{\l}kowska}}]{Szegedi-Elek2020}
{Szegedi-Elek}, E., {{\'A}brah{\'a}m}, P., {Wyrzykowski}, {\L}., {et~al.} 2020, \apj, 899, 130

\bibitem[{Toon \& Ackerman(1981)}]{toon81}
Toon, O.~B., \& Ackerman, T. 1981, Applied Optics, 20, 3657

\bibitem[{{van der Walt} {et~al.}(2011){van der Walt}, {Colbert}, \& {Varoquaux}}]{vanderWalt2011}
{van der Walt}, S., {Colbert}, S.~C., \& {Varoquaux}, G. 2011, Computing in Science and Engineering, 13, 22

\bibitem[{{Vorobyov} \& {Basu}(2006)}]{Vorobyov2006}
{Vorobyov}, E.~I., \& {Basu}, S. 2006, \apj, 650, 956

\bibitem[{{Vorobyov} \& {Basu}(2015)}]{Vorobyov2015}
---. 2015, \apj, 805, 115

\bibitem[{{Vorobyov} {et~al.}(2018){Vorobyov}, {Elbakyan}, {Plunkett}, {Dunham}, {Audard}, {Guedel}, \& {Dionatos}}]{Vorobyov2018}
{Vorobyov}, E.~I., {Elbakyan}, V.~G., {Plunkett}, A.~L., {et~al.} 2018, \aap, 613, A18

\bibitem[{{W}es {M}c{K}inney(2010)}]{mckinney2010}
{W}es {M}c{K}inney. 2010, in {P}roceedings of the 9th {P}ython in {S}cience {C}onference, ed. {S}t\'efan van~der {W}alt \& {J}arrod {M}illman, 56 -- 61

\bibitem[{{White} {et~al.}(2016){White}, {Boley}, {Hughes}, {Flaherty}, {Ford}, {Wilner}, {Corder}, \& {Payne}}]{White2016}
{White}, J.~A., {Boley}, A.~C., {Hughes}, A.~M., {et~al.} 2016, \apj, 829, 6

\bibitem[{White {et~al.}(2020)White, K{\'o}sp{\'a}l, Hughes, {\'A}brah{\'a}m, Akimkin, Banzatti, Chen, de~Miera, Dutrey, Flock, {et~al.}}]{white20}
White, J.~A., K{\'o}sp{\'a}l, {\'A}., Hughes, A., {et~al.} 2020, The Astrophysical Journal, 904, 37

\bibitem[{Woitke {et~al.}(2016)Woitke, Min, Pinte, Thi, Kamp, Rab, Anthonioz, Antonellini, Baldovin-Saavedra, Carmona, {et~al.}}]{woitke16}
Woitke, P., Min, M., Pinte, C., {et~al.} 2016, Astronomy \& Astrophysics, 586, A103

\bibitem[{{Zhu} {et~al.}(2009){Zhu}, {Hartmann}, {Gammie}, \& {McKinney}}]{Zhu2009}
{Zhu}, Z., {Hartmann}, L., {Gammie}, C., \& {McKinney}, J.~C. 2009, \apj, 701, 620

\bibitem[{Zubko {et~al.}(1996)Zubko, Kreiowski, \& Wegner}]{zubko96}
Zubko, V., Kreiowski, J., \& Wegner, W. 1996, Monthly Notices of the Royal Astronomical Society, 283, 577

\end{thebibliography}

\end{document}